\documentclass[iop,apj,tighten,revtex4]{emulateapj}
\usepackage{color}
\usepackage{graphicx}
\usepackage{natbib}
\usepackage{amssymb}
\usepackage{amsmath}
\usepackage{rotating}
\usepackage{url}

\defcitealias{2006MNRAS.369..257D}{DBK06}
\defcitealias{2014ApJS..213...10W}{Paper~I}

\shorttitle{On the Progenitors of Local Group Novae. II.}
\shortauthors{Williams et~al.}
   
\begin{document}

\title{On the Progenitors of Local Group Novae. II. The Red Giant Nova Rate of M31}

\author{S.~C. Williams$^{1}$, M.~J. Darnley$^{1}$, M.~F. Bode$^{1}$, \and A.~W. Shafter$^{2}$}
\affil{$^{1}$Astrophysics Research Institute, Liverpool John Moores University, Liverpool, L3~5RF, UK}
\affil{$^{2}$Department of Astronomy, San Diego State University, San Diego, CA 92182, USA}

\submitted{{\scriptsize Received 2015 September 26; accepted 2015 December 13}}

\journalinfo{The Astrophysical Journal, Draft version \today}

\begin{abstract}
In our preceding paper, Liverpool Telescope data of M31 novae in eruption were used to facilitate a search for their progenitor systems within archival {\it Hubble Space Telescope} ({\it HST}) data, with the aim of detecting systems with red giant secondaries (RG-novae) or luminous accretion disks.  From an input catalog of 38 spectroscopically confirmed novae with archival quiescent observations, likely progenitors were recovered for eleven systems.  Here we present the results of the subsequent statistical analysis of the original survey, including possible biases associated with the survey and the M31 nova population in general. As part of this analysis we examine the distribution of optical decline times ($t_2$) of M31 novae, how the likely bulge and disk nova distributions compare, and how the M31 $t_2$ distribution compares to that of the Milky Way. Using a detailed Monte Carlo simulation, we determine that $30^{+13}_{-10}$ percent of all M31 nova eruptions can be attributed to RG-nova systems, and at the 99~percent confidence level, $>10$ percent of all M31 novae are RG-novae. This is the first estimate of a RG-nova rate of an entire galaxy. Our results also imply that RG-novae in M31 are more likely to be associated with the M31 disk population than the bulge, indeed the results are consistent with all RG-novae residing in the disk. If this result is confirmed in other galaxies, it suggests any Type Ia supernovae that originate from RG-nova systems are more likely to be associated with younger populations, and may be rare in old stellar populations, such as early-type galaxies.
\end{abstract}


\section{Introduction}
Classical nova (CN) eruptions occur in binary systems comprising a white dwarf (WD) primary and a generally less massive (either main-sequence, sub-giant, or red giant) secondary star. The WD accretes matter from the secondary either via Roche lobe overflow or from the stellar wind. As material is accreted by the WD, the pressure and temperature at the base of the accreted envelope increase until nuclear fusion can occur, which, in the degenerate conditions found here then leads to a thermonuclear runaway (TNR; \citealp{1972ApJ...176..169S}). The energy released during the TNR causes a rapid increase in luminosity which we observe as the nova eruption \citep[see e.g.,][and references therein]{2008clno.book.....B,2010AN....331..160B,2014ASPC..490.....W}. Novae can exceed magnitudes of $M_V=-10$ \citep[see][and references therein]{2009ApJ...690.1148S}, and are one of the most energetic stellar explosions known, behind Gamma Ray Bursts and Supernovae (SNe).

M31 has a predicted nova rate of $65^{+16}_{-15}$~year$^{-1}$ \citep[hereafter DBK06]{2006MNRAS.369..257D} and over 1,000 candidates have been discovered in that galaxy over the last 100 years \citep[][see also their online catalog\footnote{\url{http://www.mpe.mpg.de/\textasciitilde m31novae/opt/m31/index.php}}]{2007A&A...465..375P,2010AN....331..187P}. Due to various limiting factors such as the seasonal gap and survey coverage (see Section~\ref{sec:discov} for a full discussion), an average of 27 novae per year have been discovered in M31 over the last ten years (excluding candidates subsequently found not to be novae)$^{1}$, which still makes it the leading environment for studying the nova population of a whole galaxy. The M31 nova population has been studied since the pioneering work of \citet{1929ApJ....69..103H}. Over the last ten years our knowledge of the properties of the M31 nova population has increased significantly, with detailed studies of the photometric, spectroscopic (e.g.\ \citealp{2011ApJ...734...12S}) and X-ray properties (e.g.\ \citealp{2014A&A...563A...2H}) being published.

The recurrent novae (RNe), by definition, have had two or more observed eruptions, unlike the single recorded eruption of each CN.  Their recurrence times range from 1 year \citep{2014A&A...563L...9D,2015A&A...580A..45D,2014A&A...563L...8H,2015A&A...580A..46H,2014ApJ...786...61T}, or even as short as six months \citep{2015A&A...582L...8H}, up to 100 years \citep[an observational upper limit;][]{2010ApJS..187..275S}, with a theoretical lower limit of two months \citep{2014ApJ...793..136K,2015ApJ...808...52K}. Of the ten confirmed Galactic RNe, half harbor red giant secondaries (RG-novae), with three containing sub-giant secondaries (SG-novae) and only two with main-sequence companions \citep[see][and references therein]{2012ApJ...746...61D}. The CN population was thought to be dominated by systems containing a main-sequence secondary (MS-novae). However, recent work on old Galactic novae suggests a significant proportion of the CN population may be RG-nova systems \citep{2014ApJ...788..164P}.  Based on a reanalysis of archival data, \citet{2015ApJS..216...34S} have shown that as many as one in three nova eruptions observed in M31 may arise from RNe (a large proportion of which are expected to be RG-novae if the Galactic sample is replicated in M31).  The subsequent analysis by \citet{2015ApJS..216...34S} predicts, even at such an elevated level, RNe are unlikely to provide the dominant channel for Type Ia SNe (SNe~Ia). 

RG-novae differ observationally from the SG and MS-novae in a number of ways, both at quiescence and during eruption. Mass transfer in RG-nova systems is typically wind driven (as opposed to Roche lobe overflow in the case of MS- and SG-novae) and as such, RG-novae are generally expected to have dense circumbinary media \citep[see, for example,][]{2010Ap&SS.327..207W}, which can interact with the ejecta and produce narrow emission line features. Such narrow components have been observed in early spectra of RS~Ophiuchi \citep{2009A&A...505..287I} and V407~Cygni \citep{2011MNRAS.410L..52M,2011A&A...527A..98S}. However, this may not be exclusive to RG-novae, as early narrow emission was also seen in the recent SG-nova candidate V2944~Ophiuchi \citep{2016MNRAS.455L..57M}. This interaction between the red giant wind and the ejecta can also produce shocks observable in the X-ray \citep{1967BAN....19..227P,2006ApJ...652..629B,2008ASPC..401..269B}. While the peak luminosity of RG, SG and MS-novae appear comparable, as the optical quiescent spectrum of RG-novae tends to be dominated by the luminous companion, the amplitude of the eruption appears smaller (typically $\sim5-6$\,magnitudes). The orbital periods of RG-novae are much longer, often over 100~days, compared to the timescale of hours in MS-novae, or days for SG-novae. These observational features of RG-novae in eruption and quiescence can all be used to indicate the presence of a giant companion in Galactic novae. However, at the distance of M31, such observations are largely beyond the capabilities of current facilities. At present, locating the quiescent nova systems (and thus eruption amplitude) is the only feasible way to identify a significant sample of M31 RG-novae.

It is generally understood that SNe~Ia are caused by the thermonuclear explosion of a carbon-oxygen (CO) WD as it surpasses the Chandrasekhar mass (e.g.\ \citealp{2000ARA&A..38..191H}). Oxygen-neon WDs are not thought to produce SNe Ia \citep{1996ApJ...459..701G}, although it may be possible for some sub-Chandrasekhar mass SNe Ia \citep{2015A&A...580A.118M}. The exact nature of their progenitor systems and the mechanism of mass transfer remain unclear. The two commonly proposed progenitor pathways are the single-degenerate and double-degenerate channels. The double-degenerate scenario involves the merger of two WDs, whereas the single-degenerate scenario involves a WD accreting non-degenerate material from a companion star and increasing in mass until the SN explosion occurs.

RG-novae are one of the single-degenerate SN Ia progenitor candidates. This scenario begins with the higher-mass star in a binary system evolving to leave behind a CO WD. The WD then begins significant accretion from the companion when the secondary starts to evolve towards a red giant, producing nova eruptions. If the WD has a net increase in mass over the nova cycle, it would be expected to eventually explode as a SNe Ia. A key parameter in assessing the likelihood of novae producing SNe Ia is the degree of mixing between the WD core and the accreted envelope prior to the nova eruption. If there is no mixing between the two, simulations predict that the WDs can increase in mass over time, even for low-mass WDs \citep{2012BASI...40..419S}. The SN Ia PTF~11kx has been suggested as originating from a RG-nova system \citep{2012Sci...337..942D}. Several other SNe Ia also appear to show interactions with a hydrogen-rich circumstellar medium (known as SNe Ia-CSM; \citealp{2013ApJS..207....3S}). It is foreseeable at the present time that there are multiple pathways leading to a SN~Ia explosion, and as such, it is important to fully understand the RG-nova population to be able to determine their contribution to the overall SNe~Ia population.

\begin{figure}
\includegraphics[width=\columnwidth]{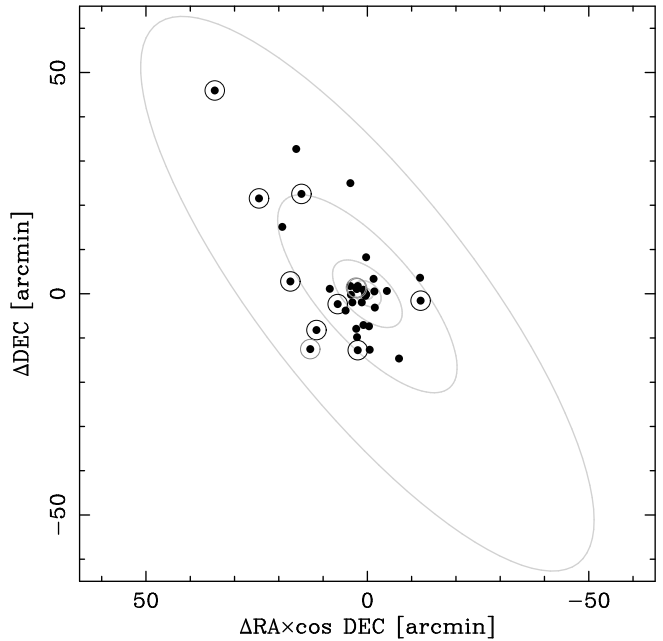}
\caption{The spatial distribution of the 38 novae from the survey for M31 nova progenitors published in \citetalias{2014ApJS..213...10W}. The black points represent the 38 novae in the survey, with the eleven with likely resolved progenitors further circled. The eight RG-nova candidates used in this analysis, those found in ACS/WFC data from LT eruption images, are circled in black, with the other three circled in dark gray (see Section~\ref{sec:two}). The light gray ellipses represent isophotes from the surface photometry of M31 from \citet{1987AJ.....94..306K}.\label{spatial}}
\end{figure}

In a preceding paper, \citet[][hereafter Paper~I]{2014ApJS..213...10W}, the results of the first extragalactic survey of nova progenitor systems was presented.  This survey was based on an input catalog of 38 spectroscopically confirmed M31 novae, and employed astrometry from Liverpool Telescope \citep[LT;][]{2004SPIE.5489..679S} observations of the eruptions to attempt to locate a corresponding quiescent system in archival {\it Hubble Space Telescope} ({\it HST}) images.  The survey found eleven of the 38 systems (29 percent) were coincident with a quiescent source in the archival {\it HST} data (one of which, M31N~2007-12b, had been previously identified by \citealp{2009ApJ...705.1056B}). The positions of the 38 novae used for the input catalog in \citetalias{2014ApJS..213...10W} are shown in Figure~\ref{spatial}, and the eleven systems with resolved progenitors further circled.

In this follow-up paper, we analyze the biases and statistics of the catalog, to explore what the results for these 38 novae tell us about the M31 nova population as a whole. Determining the proportion of novae with evolved secondaries is important for understanding the contribution they may make to the SN Ia progenitor population.

\section{The Model}\label{sec:two}
If we look back at the catalog published in \citetalias{2014ApJS..213...10W}, in order to be included in the progenitor search any given nova had to:

\begin{enumerate}
\item Produce an eruption between August 2006 and February 2013; the timeframe of the \citetalias{2014ApJS..213...10W} survey.
\item Be discovered. While this may be stating the obvious, it is important to consider any effect discovery biases may have had on our results.
\item Be spectroscopically confirmed. In order to be certain of the nova nature of the transient.
\item Have had astrometric data taken while still in eruption.
\item Have had spatially coincident {\it HST} data taken while at quiescence.
\end{enumerate}

The \citetalias{2014ApJS..213...10W} catalog of eleven novae with resolved progenitors was produced from an input catalog of 38 spectroscopically confirmed novae, that had either {\it HST} WFPC2 or ACS/WFC quiescent data.  However, as discussed in \citetalias{2014ApJS..213...10W}, the WFPC2 data are typically not deep enough to fully sample the red giant branch in M31. So, while we can generally believe a positive detection of a nova progenitor system from WFPC2 data, a non-detection is generally uninformative. To counter this effect, we limit ourselves to just ACS/WFC data for this analysis; to include the WFPC2 data in this analysis would require the inclusion of a model of the M31 stellar luminosity distribution, which is beyond the scope of this work.  This decision truncates the input catalog to 33 novae and essentially eliminates the main source of potential false negatives for the RG-nova population, leaving us only concerned with the false positives, as is discussed in Section~\ref{self-const}. In this analysis, we only consider RG-novae for which the progenitor was found using LT eruption images. The few with additional {\it HST} observations taken when in eruption, would have a better chance of progenitor recovery if they were associated with a quiescent source, due to the higher astrometric accuracy of the {\it HST} data compared to that of the LT. If these were included as progenitor detections the considerations of Section~\ref{sec:dataan} would not accurately reflect the catalog. For the purposes of the analysis described in this paper, this therefore gives us eight of 33 novae found to be likely RG-novae.

Throughout this analysis we assume all eight of the \citetalias{2014ApJS..213...10W} progenitor systems are either RG-novae or simply due to chance alignments (i.e.\ we assume only novae with red giant secondaries are detectable in the {\it HST} data in quiescence).  We present the evidence on which we base this assumption, and the implications should it be invalid, in Section~\ref{nature}.  This assumption is made, partly for ease of description, based on that evidence.

To use the results of our survey of individual novae to understand the entire nova population of M31, we need to consider how typical, and statistically significant, our sample was. In order to consider the potential effects of the biases associated with each of the above steps, we simulate the M31 nova population, which is then used in a Monte Carlo model to conduct a full statistical analysis.

\subsection{Simulating The Novae} \label{simul}
One of the primary inputs to a Monte Carlo simulation of the \citetalias{2014ApJS..213...10W} survey is a detailed description of the nova population of M31, its spatial distribution, the range of speed class and luminosity, and the observability of each simulated eruption.  Here, we describe our modeling of the nova population and its observability.

\subsubsection{Nova Spatial Distribution} \label{sec:spa}
A number of surveys of the M31 nova population have indicated that M31 novae effectively follow the bulge light \citep[see for example,][\citetalias{2006MNRAS.369..257D} and references therein]{2001ApJ...563..749S}.  \citetalias{2006MNRAS.369..257D} confirmed that the majority of M31 novae follow the bulge light, but they also showed there was a smaller but still significant population that followed the disk light.  Here, the spatial distribution of M31 novae predicted by the \citetalias{2006MNRAS.369..257D} model was used to simulate the positions of the seeded novae. This model, which is described in detail by \citet{2005PhDT.........2D} and \citetalias{2006MNRAS.369..257D} (see their Equation~22), assumes there are two separate populations of novae in M31, one following the disk light and one following the bulge light.  The model is described by

\begin{equation}
\Psi_{i} = \frac{\theta f_{i}^{d} + f_{i}^{b}}{\theta\sum_{j} f_{j}^{d} + \sum_{j} f_{j}^{b}},\label{DBK06model}
\end{equation}

\noindent where $\Psi$ is the probability of a nova erupting at a given location, $\theta$ is the ratio of disk and bulge nova eruption rates per unit ($r'$-band) flux, $f^{d}$ and $f^{b}$ are the disk and bulge contributions to the ($r'$-band) flux at that location, respectively, the terms in the denominator are summed over all locations in M31 normalizing the distribution, such that $\sum_{i}\Psi_{i}=1$. This equation was first employed to investigate the spatial distribution of the M31 nova population by \citet{1987ApJ...318..520C}.

The relative contributions of the two populations to the overall nova distribution was originally calculated by comparing the disk and bulge flux of M31 to the novae observed in the POINT--AGAPE survey of M31 \citep[see][]{2004MNRAS.353..571D}. This model relies on a single parameter, $\theta$, that \citetalias{2006MNRAS.369..257D} constrain to $\theta=0.18^{+0.24}_{-0.10}$ ($1\sigma$ limits).  Throughout this paper we assume a value of $\theta=0.18$, which leads to a nova rate of $65\pm15$~year$^{-1}$, with a contribution of $38\pm8$~year$^{-1}$ and $27\pm6$~year$^{-1}$ from the bulge and disk populations\footnote{The uncertainties quoted for these rates differ from those given in DKB06 as here we have removed any contribution from the uncertainty on $\theta$.}, respectively (however, our analysis is not directly sensitive to the overall rate, simply the bulge to disk ratio).  We discuss the sensitivity of our subsequent analysis to the choice of $\theta$ in Section~\ref{paramsens}.

\subsubsection{Nova Observability} \label{sec:t2dist}
To take account of the observability of a given nova eruption in M31 we must consider the surface brightness of the galaxy at the position of that nova, the peak luminosity, and the decline rate of that nova.

The rate that novae fade in the optical is often classified by the time it takes them to fade by two magnitudes from peak, $t_{2}$.  Previous work on both Galactic and M31 novae has shown evidence to support slower fading novae being more likely to be associated with the older bulge population of a galaxy, whereas novae in the disk tend to be faster \citep{1990LNP...369...34D,1992A&A...266..232D,2011ApJ...734...12S,2012ApJ...752..156S}.  

\begin{figure*}
\includegraphics[width=0.49\textwidth]{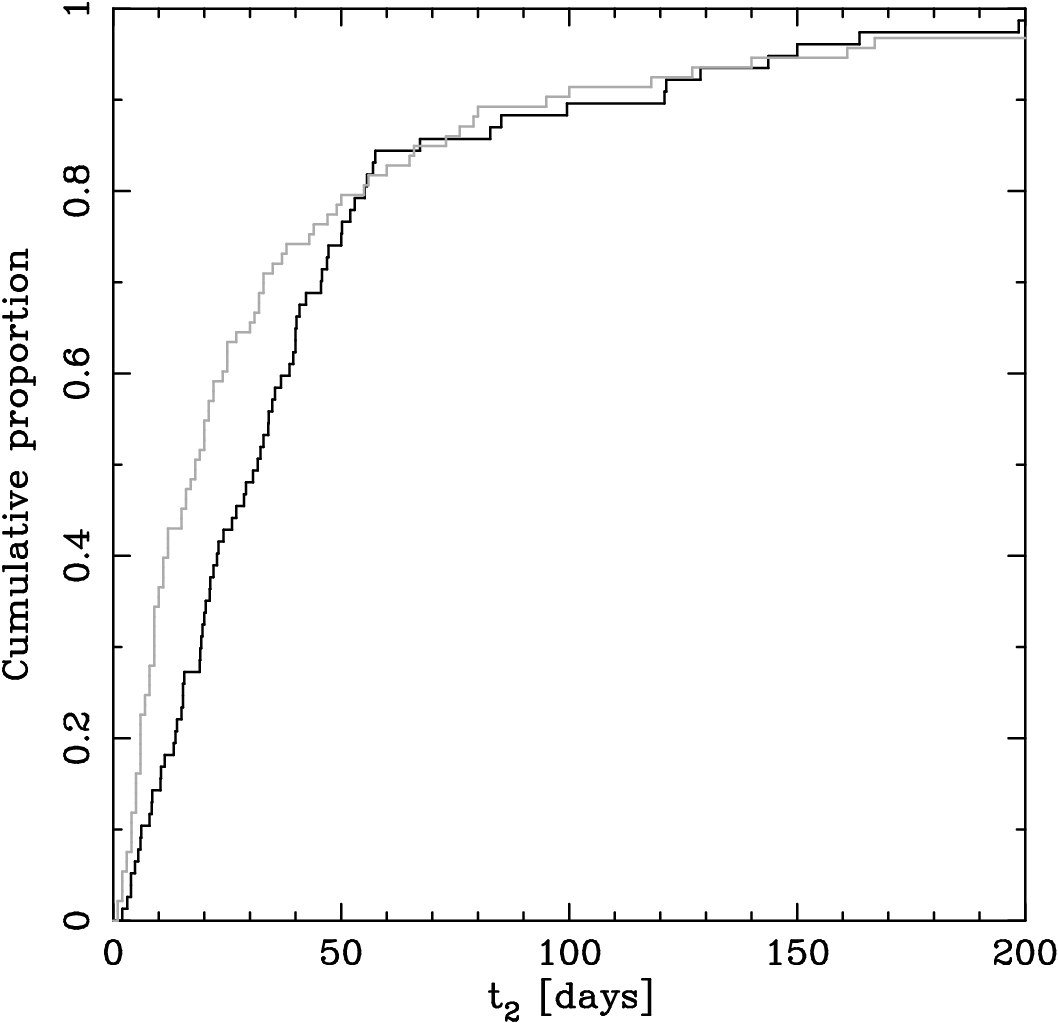}
\hfill
\includegraphics[width=0.49\textwidth]{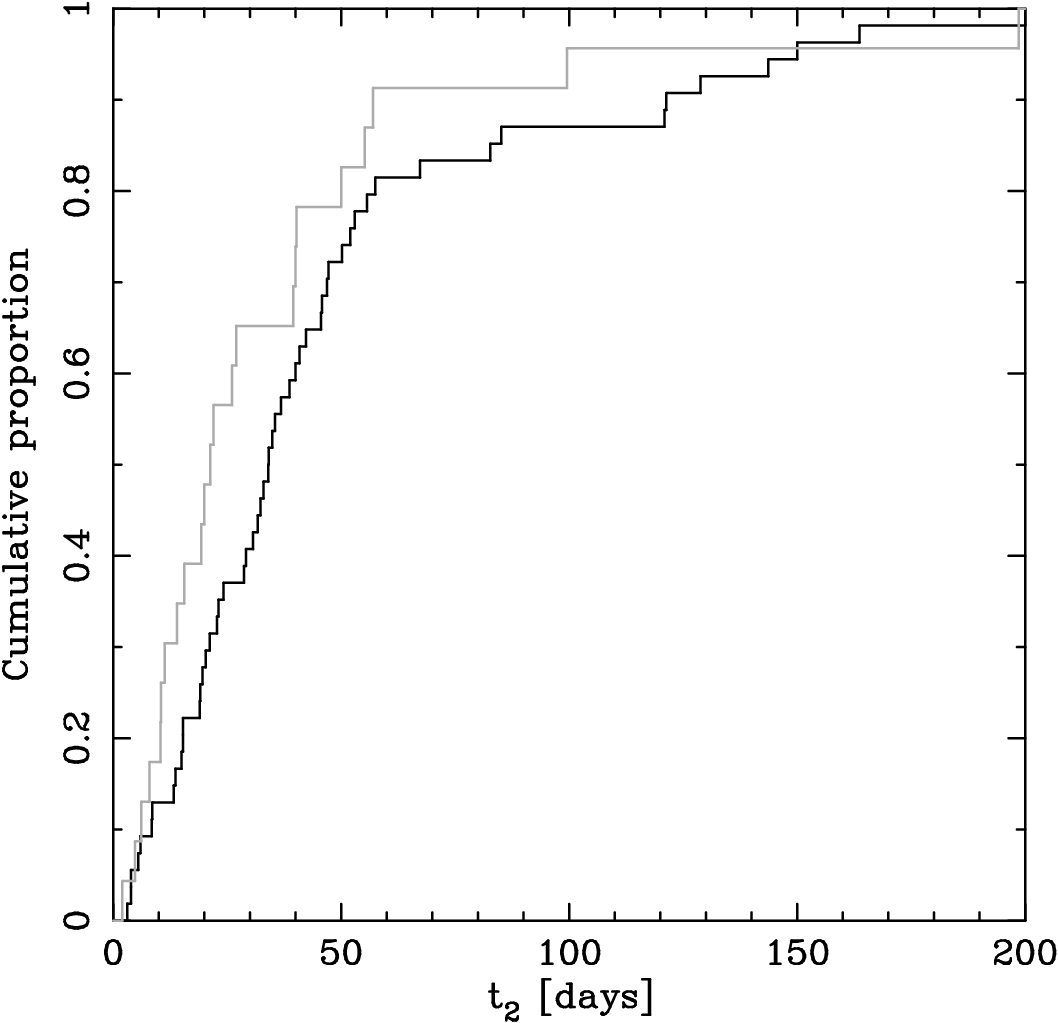}
\caption{The cumulative $t_2$ distribution of novae in M31, using $t_2$ values from \citet{2005PhDT.........2D}, \citet{2011ApJ...727...50S,2011ApJ...734...12S}, \citet{2012ApJ...752..133C}, \citetalias{2014ApJS..213...10W} and \citet{williamsPhD}. {\bf Left:} The cumulative $t_2$ distribution of all novae in M31 (black line) compared the Galactic novae from \citet[][gray line]{2010AJ....140...34S}. {\bf Right:} The black line represents the $t_2$ distribution of novae occurring within $15^{\prime}$ of the center of M31 and the gray line represents the $t_2$ distribution of novae occurring further than $15^{\prime}$ from the center. The only M31 $t_2$ value not seen on these plots is M31N~2008-07a (see Section~\ref{sec:t2dist}).\label{t2dist}}
\end{figure*}

To assess the $t_{2}$ distribution of M31 novae, we compiled M31 nova $t_{2}$ data from \citetalias{2006MNRAS.369..257D} and spectroscopically confirmed eruptions from \citet{2011ApJ...727...50S,2011ApJ...734...12S}, \citet{2012ApJ...752..133C}, \citetalias{2014ApJS..213...10W} and \citet{williamsPhD}, totalling 77 individual systems. For comparison, we used Galactic $t_{2}$ values of 93 novae from \citet{2010AJ....140...34S}. The cumulative $t_{2}$ distribution of M31 novae, along with a comparison to Galactic novae, is shown in the left hand panel of Figure~\ref{t2dist}. 

Due in part to the inclination of M31, it is difficult to unambiguously determine if a given M31 nova is part of the disk or bulge. Therefore, to investigate any difference between the $t_{2}$ distributions of the M31 bulge and disk populations we use simply the radius from the center of M31 as a discriminant.  The \citetalias{2006MNRAS.369..257D} model predicts that the cross-over point between bulge nova dominance and disk nova dominance occurs at a radius of $\sim15^{\prime}$.  Novae within this radius are $>50$~percent likely to be bulge novae, outside they are $>50$~percent likely to be disk novae.  A comparison between the cumulative $t_{2}$ distributions of M31 novae within and outside $15^{\prime}$ is shown in the right hand panel of Figure~\ref{t2dist}. This plot indicates that a higher proportion of fast novae are found beyond $15^{\prime}$, in the disk nova dominated portion of M31.  An \citet{anderson1952} test (AD-test) comparing the M31 `bulge' and `disk' $t_{2}$ distributions suggests the probability that they are drawn from the same population is 0.18\footnote{The AD-test was used in preference over the \citet{kolmogorov33}--\citet{smirnov1948} test (KS-test) as the sensitivity of the KS-test varies across the distribution, being least at the extremes of any distribution \citep[see, for example,][]{doi:10.1080/01621459.1974.10480196}.  In all cases employed here, the KS- and AD-tests give similar results.}.  While not conclusive, this supports the earlier works that concluded disk novae are more likely to fade faster than those occurring in the bulge.

There is a sharp break in the M31 $t_{2}$ distributions that occurs between $t_{2}=50$ and $t_{2}=60$~days, seen particularly well in the overall M31 distribution.  The cause of this break is unknown, as it is not seen in the Galactic data (see Figure~\ref{t2dist}) which is generally smooth.  The M31 seasonal observing gap may have some effect on the ability to measure the decline rate, as slower novae must be followed for longer in order to determine $t_{2}$, but this would not be expected to produce such a pronounced break.  Additionally, there appears to be a higher proportion of fast novae in the Galaxy than in M31.  The M31 sample is likely to be more representative of the overall nova population than the Galactic sample, as we effectively see the whole of M31, but only a section of our Galaxy. This bias is probably even stronger in the older data (which \citealp{2010AJ....140...34S} include) where the ability to detect and follow fainter (hence more likely to be slow or distant) novae was diminished.  An AD-test comparing the overall M31 $t_{2}$ cumulative distribution with that from the Galaxy indicates a probability of 0.006 that they are drawn from the same population.  However, an AD-test between the Galactic distribution and novae $>15^{\prime}$ from the center of M31 (the M31 `disk' novae) reveals a probability of 0.52 that they are drawn from the same population.  From this result we can infer the observed Galactic $t_{2}$ distribution may be dominated by disk novae, due to its similarity to the M31 disk $t_{2}$ distribution.  In addition, we must conclude that the M31 bulge novae population, which dominates the observed M31 novae, is not similar to the observed Galactic population (on which our general understanding of novae is based).  Finally, for completeness, an AD-test between the Galactic $t_{2}$ distribution and the M31 `bulge' novae indicates a probability of 0.002 that they are drawn from the same population.

Although the evidence presented above indicates that the $t_{2}$ distributions in the M31 bulge and disk may differ, when simulating the M31 nova population we simply refer to the overall M31 distribution (see left panel of Figure~\ref{t2dist}). The $t_{2}$ cumulative distribution is modeled simply with three straight lines (excluding nova M31N~2008-07a, as its decline rate is unusually slow and also highly uncertain at $t_2=410.0\pm51.3$~days; \citealp{2011ApJ...734...12S}), one fit to data where $t_{2}<30$~days, one where $30<t_{2}<60$~days and the other $t_{2}>60$~days.  When modeling the overall nova population, each nova is randomly assigned a decline time based on this modeled cumulative distribution.  It is also worth noting the results presented in Section~\ref{themodel} and beyond do not vary significantly if we use separate ``disk" and ``bulge" $t_{2}$ distributions. We assume the $t_{2}$ distribution of RG-novae is the same as that for the general population.

Finally, each simulated nova was also assigned a peak magnitude based on their $t_{2}$ time. This was performed by using the maximum magnitude--rate of decline relationship ({\it R}-band) for M31 novae from \citet{2011ApJ...734...12S}, 

\begin{equation}
M_{R}=-10.89\pm0.12+\left(2.08\pm0.08\right)\log{t_{2}},\label{mmrd_eq}
\end{equation}

\noindent where overall, the slower novae tend to be less luminous.  The observability of each simulated eruption could then be computed as the time for which a nova was visible above the M31 surface brightness, given its spatial position, decline time, and peak luminosity.

\subsection{Simulating the Data Analysis} \label{sec:dataan}
\subsubsection{Astrometry}
The \citetalias{2014ApJS..213...10W} survey was fundamentally an astrometric survey, and to fully simulate the data analysis we need to assess the impact of any astrometric uncertainties.  Due to the resolved stellar density in the {\it HST} data the probability of a chance alignment between an erupting nova and an unrelated resolved star is relatively high.  To counter this effect in \citetalias{2014ApJS..213...10W}, we introduced a requirement that the chance alignment probability be $\leq5$~percent for a positive progenitor recovery \citep[we note this is much more conservative than the criterion used for a similar statistic by][]{2015ApJS..216...34S}.  With such a strict constraint, it is therefore possible that a nova with a resolvable progenitor system would not pass this test due to relatively large astrometric uncertainties from the eruption data, and vice versa.  

As described in Section~\ref{simul}, each seeded nova was given, among other parameters, a position in M31.  We simulated the astrometric uncertainties by introducing a Gaussian distributed random offset between the position of the seeded erupting nova and its progenitor system.  This offset was based upon the median astrometric uncertainties calculated during the \citetalias{2014ApJS..213...10W} survey, $\sigma=0^{\prime\prime}\!\!.0385$.

\subsubsection{Chance Alignments} \label{sec:chance}
\begin{figure}
\includegraphics[width=\columnwidth]{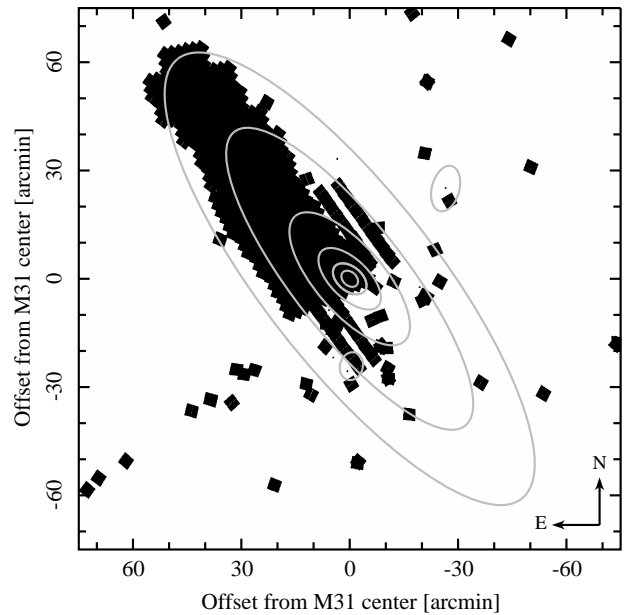}
\caption{The ACS {\it HST} coverage of M31. Covered regions appear in black, with regions with no coverage appearing white. The gray ellipses represent isophotes from the surface photometry of M31 from \citet{1987AJ.....94..306K}, with M32 and NGC~205 also indicated.\label{hstcov}}
\end{figure}

\begin{figure}
\includegraphics[width=\columnwidth]{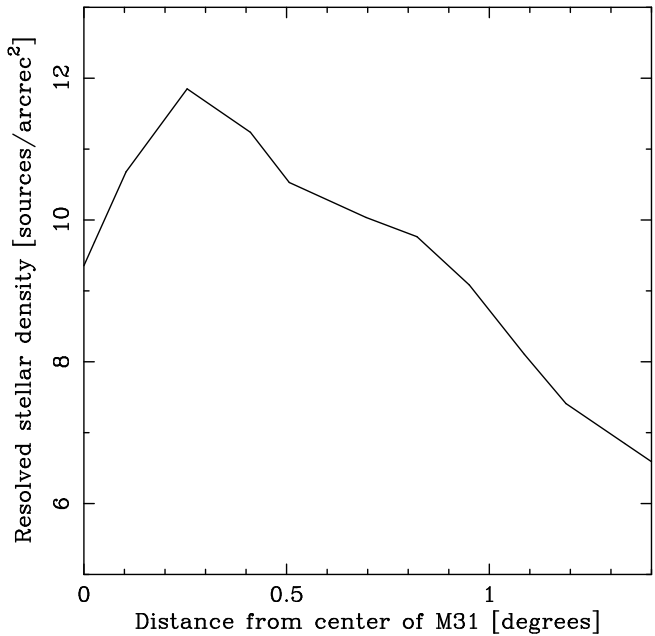}
\caption{The density of resolved stellar sources in the M31 field as a function of distance from the center of the galaxy, computed along the north-eastern axis of M31.\label{fig:dens}}
\end{figure}

One of the main difficulties encountered during the \citetalias{2014ApJS..213...10W} survey was the high density of resolved sources in the M31 field, leading to the possibility of chance alignment with an erupting nova.   To understand the effects of chance alignments, and its dependence upon position in M31, we created a simple model of the M31 source density.  We calculated the spatial density of resolved stellar sources in the ACS/WFC {\it HST} data at eleven set intervals along the (north-eastern) semi-major axis of M31. Photometry of all sources within a $\sim2$ square-arcmin region at these intervals was performed on the ACS/WFC archival data (see Section~\ref{HST} for further details) using DOLPHOT\footnote{\url{http://americano.dolphinsim.com/dolphot}} \citep[v1.1;][]{2000PASP..112.1383D}, following the prescription described in \citetalias{2014ApJS..213...10W}. The north-eastern axis of M31 was chosen as the {\it HST} archive is essentially complete along this axis (see Figure~\ref{hstcov} and the discussion in Section~\ref{HST}).  The resulting stellar density distribution is shown in Figure~\ref{fig:dens}.  Here the stellar density can be seen to increase with decreasing radius, until $\sim0^{\circ}\!.25$ from the center where the rapidly increasing surface brightness begins to hinder the detection of the fainter sources.  To include this model in our simulation we assume that the resolved stellar density varies smoothly and employ simple linear interpolation between data points.  We also assume a rotational symmetry to this distribution, and treat M31 as a flat disk inclined so that the minor/major axis ratio $=0.31$, with the density being uniform at any point in the disk at a given radius.  We can therefore estimate the resolved stellar density at any point within M31 and use this information to compute the significance of any progenitor recovery.

\subsubsection{Progenitor Recovery Significance} \label{sec:sig}
To compute the significance of any progenitors recovered in the \citetalias{2014ApJS..213...10W} survey we originally employed a time-consuming Monte Carlo technique for each of the 38 systems.  However, as we expected to test a significantly larger number of systems in this simulation, we instead turned to a general analytical solution to this problem.

Considering {\it N} sources within an area of {\it A} giving a spatial density of {\it n}, the probability of a source being within an annulus of radii $r$ and $r+dr$ is $2 {\pi} r n dr$ and the probability of a source being outside a circle of radius {\it r} is given by

\begin{equation}
1 - \frac{\pi r^{2}}{A}.
\end{equation}

\noindent Therefore the probability of a source being within an annulus of radii $r$ and $r+dr$, with that source being the closest to the center of the annulus is given by

\begin{equation}
P_{N}(r)dr = 2 \pi nrdr \left( 1 - \frac{\pi r^{2}}{A} \right)^{N-1}. 
\end{equation}

\noindent Taking $N \rightarrow \infty$, the probability of a source being within a radius, {\it r} is

\begin{equation}
\int_{0}^{r} P(r)dr = 1 - e^{-\pi nr^{2}}. 
\end{equation}

\noindent If the probability of chance alignment of $\leq5$ percent is used as the criterion for likely progenitor recovery, as in \citetalias{2014ApJS..213...10W}, then the maximum distance between a source and the position of the nova so that the probability of chance a alignment $\leq5$ percent is achieved is given by

\begin{equation}
r =\sqrt{-\frac{\ln{0.95}}{\pi n}}. \label{equ:signif}
\end{equation}

The criterion shown in \noindent Equation~(\ref{equ:signif}) was therefore used to check if the closest progenitor candidate source to a given nova eruption would constitute a significant detection (i.e.\ $\leq5$~percent probability of being a chance alignment).

\subsection{Survey Biases} \label{sec:biases}
We now consider the potential biases in the \citetalias{2014ApJS..213...10W} survey, so we can account for them in our simulation.

\subsubsection{Probability of Detection} \label{sec:discov}
Many M31 novae are found by dedicated `professional' surveys \citep[e.g., PTF/iPTF;][]{2009PASP..121.1395L,2012ApJ...752..133C}, however, a large number are discovered by `amateur' astronomers. There are numerous biases associated with the discovery of the novae both spatial and temporal, with, for example, the observed  spatial nova distribution (the detection distribution) not following the underlying eruption distribution \citep[see e.g.,][]{2015ApJS..216...34S}.  

The spatial detection efficiency of novae in M31 is a strong function of time, particularly when concerted M31 nova surveys were in progress.  Therefore, we addressed any spatial detection biases by directly comparing the observed distribution of nova candidates during the time of the \citetalias{2014ApJS..213...10W} survey (August 2006 -- February 2013) to the distribution predicted by the  DKB06 model.  This analysis suggested that novae were more likely to be discovered if they erupted between $1^{\prime}$ and $15^{\prime}$ from the center of M31, novae towards the center of the galaxy and those in the outer disk were less likely to be discovered.  If the probability of discovery is normalized to the discovery chance of novae occurring between $1^{\prime}$ and $15^{\prime}$, we find the relative probability of novae being discovered within $1^{\prime}$ of the center is only $\sim0.27$, likely due to the high surface brightness making it difficult to observe such novae. For novae erupting between $15^{\prime}$ and $70^{\prime}$ and $>70^{\prime}$ from the center, the relative discovery probabilities are $\sim0.67$ and $\sim0.40$ respectively. This is likely due to the more remote regions of M31 not being covered by some surveys.  To account for any spatial discovery bias, the \citetalias{2006MNRAS.369..257D} model is convolved with the above detection likelihoods to determine an M31 nova spatial detection model. 

The main problem in observing novae in M31 has always been the temporal coverage, as M31 is difficult, if not impossible, to observe for several months of each year.  In addition, the temporal availability of observing effort has major impact on the discovery efficiency, as is shown by the data presented in Table~\ref{tab:monthrate}.  We used these data to estimate and simulate the relative discovery efficiency of M31 novae as a function of time.

\begin{deluxetable}{lcr}
\tablewidth{\columnwidth}
\tablecaption{M31 nova candidates discovered in each calendar month / year from the
start of 2006 to the end of 2012.\label{tab:monthrate}}
\tablehead{
\colhead{Month / year} & \colhead{Novae discovered} & \colhead{$\Delta\mathrm{RA}$\tablenotemark{1}}
}
\startdata
Jan & 15 & 5$^{\mathrm{h}}$ \\
Feb & 17 & 3$^{\mathrm{h}}$ \\
Mar & 9 & 1$^{\mathrm{h}}$ \\
Apr & 2 & 1$^{\mathrm{h}}$ \\
May & 15 & 3$^{\mathrm{h}}$ \\
Jun & 26 & 5$^{\mathrm{h}}$ \\
Jul & 15 & 7$^{\mathrm{h}}$ \\
Aug & 20 & 9$^{\mathrm{h}}$ \\
Sep & 14 & 11$^{\mathrm{h}}$ \\
Oct & 22 & 11$^{\mathrm{h}}$ \\
Nov & 24 & 9$^{\mathrm{h}}$ \\
Dec & 19 & 7$^{\mathrm{h}}$ \\
\hline
2006 & 19 & \nodata \\
2007 & 32 & \nodata \\
2008 & 31 & \nodata \\
2009 & 22 & \nodata \\
2010 & 28 & \nodata \\
2011 & 31 & \nodata \\
2012 & 28 & \nodata
\enddata
\tablenotetext{1}{The average difference in Right Ascension between M31 and the Sun in each month.}
\end{deluxetable}

\subsubsection{Astrometric Observations}
One of the keys to the analysis in \citetalias{2014ApJS..213...10W} was obtaining high-precision astrometry of each nova eruption.  This necessitated rapid follow-up observations of each eruption, the robotic nature of the LT made this an ideal facility, and all of our astrometric observations were undertaken by the LT.  The main hindrance in obtaining the LT observations was the M31 seasonal gap. The LT cannot observe below an altitude of $25^{\circ}$, and is unable to observe M31 during the majority of March, April and May.  Telescope technical downtime, planned maintenance, weather, and observing conditions also affected the ability to obtain astrometry throughout the survey.  For example, the longest period where the LT was unable to observe during the \citetalias{2014ApJS..213...10W} survey was 17 consecutive nights.  To fully assess the effect of any survey biases introduced by the LT, our simulations must effectively reproduce the LT's availability during the survey.  By coupling the M31 observing constraints of the the LT with detailed observing logs\footnote{Obtained from \url{http://telescope.livjm.ac.uk}} we can create a simple model of LT availability.  Here, we must also take account of the observability (see Section~\ref{sec:t2dist}) of each simulated nova.  For example, if a simulated nova erupted during LT downtime, we must determine if, based on the brightness, decline time, and position in M31, that nova would still be observable at the next LT observing opportunity. 

\subsubsection{Spectroscopic Observations}
The spectroscopic confirmation of the novae in the \citetalias{2014ApJS..213...10W} catalog was also crucial.  The misidentification of long period variables, particularly Miras, as M31 novae \citep[see, for example,][]{2004MNRAS.353..571D} could have formed a pool of potential false positives within our catalog.  Such systems are only visible briefly, from most ground-based facilities, when they brighten above the M31 surface luminosity.  By their very nature, as luminous, evolved stars, they will appear resolved within deep archival data. Other possible contaminants where a progenitor may be resolvable include luminous red novae in M31 (see e.g.\ \citealp{2015ApJ...805L..18W}) or foreground dwarf novae (see e.g.\ \citealp{2008ATel.1867....1K}). Although the majority of our spectral observations were provided by the Hobby Eberly Telescope \citep[HET; e.g.][see their Table~4]{2011ApJ...734...12S}, to maximize the size of the input catalog, we relied upon confirmation spectra taken by any telescope.  The practical difficulties affecting the spectroscopic observations are the same as encountered by the astrometric observations; however, we now need to assess these across numerous telescopes.  Rather than creating detailed simulations of the observing conditions at numerous telescopes over a 6.5~year period (as we did for the LT), we based our model of the availability of spectroscopic observations on the statistical properties of the spectra that were taken.  For the temporal availability of confirmation spectra see the data presented in Table~\ref{tab:specconf}.  Spectroscopic observations in M31, particularly towards the center of the bulge, can additionally suffer from acquisition problems, mainly to due the lack of available field stars.  Indeed, a detailed look at the nova candidates occurring during the time of the \citetalias{2014ApJS..213...10W} survey reveals that a smaller proportion of the novae occurring near the center of M31 had available spectra. If the probability of spectroscopic confirmation is normalized to only consider the spatial bias (i.e.\ for novae in the outer regions, normalized probability of confirmation now $=1$), we find that nova eruptions in the central $2^{\prime}$ of M31 have a relative probability of spectroscopic confirmation of 0.67, and novae occurring $2^{\prime}-5^{\prime}$ from the center have a relative confirmation probability of 0.85. For novae $>5^{\prime}$ from the center, the probability of spectroscopic confirmation appears to be largely uniform (so relative probability $=1$).

\begin{deluxetable}{lr}
\tablewidth{\columnwidth}
\tablecaption{Proportion of M31 nova candidates with confirmation spectroscopy in each calendar month from the
start of Aug 2006 to the end of Feb 2013\label{tab:specconf}}
\tablehead{
\colhead{Month} & \colhead{Proportion confirmed}
}
\startdata
Jan & 0.44 \\
Feb & 0.28 \\
Mar & 0 \\
Apr & 0 \\
May & 0.21 \\
Jun & 0.63 \\
Jul & 0.80 \\
Aug & 0.76 \\
Sep & 0.85 \\
Oct & 0.90 \\
Nov & 0.75 \\
Dec & 0.72
\enddata
\end{deluxetable}

\subsubsection{Archival {\it HST} Coverage}\label{HST}
As not all regions of M31 have pre-existing {\it HST} coverage, it is effectively the availability of these data that limits the spatial extent of our analysis.  As such, we must consider how the availability of {\it HST} data has affected the results of the \citetalias{2014ApJS..213...10W} survey. The FITS file headers of all {\it HST} imaging data within the vicinity of M31 were interrogated using the Starview program\footnote{\url{http://starview.stsci.edu}} hosted by the Space Telescope Science Institute (STScI). The spatial coordinates of each {\it HST} observation, the position angle, instrument, filter, exposure time, and data and time were retrieved for each observation.  These data allowed us to compile a full picture of the archival data availability, particularly its growth over time.  In this analysis, only ACS data were considered, as this covered 33 of the 38 novae in our survey, and the red-giant branch is generally resolvable in the ACS data, which is not the case for that taken using WFPC2. The ACS (both WFC and HRC) {\it HST} coverage of the M31 field (as of January 2014) is shown in Figure~\ref{hstcov}.  The majority of the ACS/WFC coverage of M31 was provided by the Panchromatic Hubble Andromeda Treasury survey \citep[PHATs; see, for example,][]{2012ApJS..200...18D}.  As is shown graphically in Figure~\ref{hstcov} the majority of the M31 light in the north-east half of the galaxy has coverage, whereas the south-west coverage is sparse at best.

Some of the novae in the \citetalias{2014ApJS..213...10W} catalog did not have any pre-eruption {\it HST} data available, so, in these cases, post-eruption data were used. Post-eruption data were only used if these data were taken a suitability long time after the eruption, or if the nova could be determined to be in quiescence by the availability of multiple epochs of post-eruption data (see also \citealp{williamsPhD}), in \citetalias{2014ApJS..213...10W} these novae were treated on a case by case basis (see Section~\ref{nature} for further discussion of individual systems).

\begin{figure}
\includegraphics[width=\columnwidth]{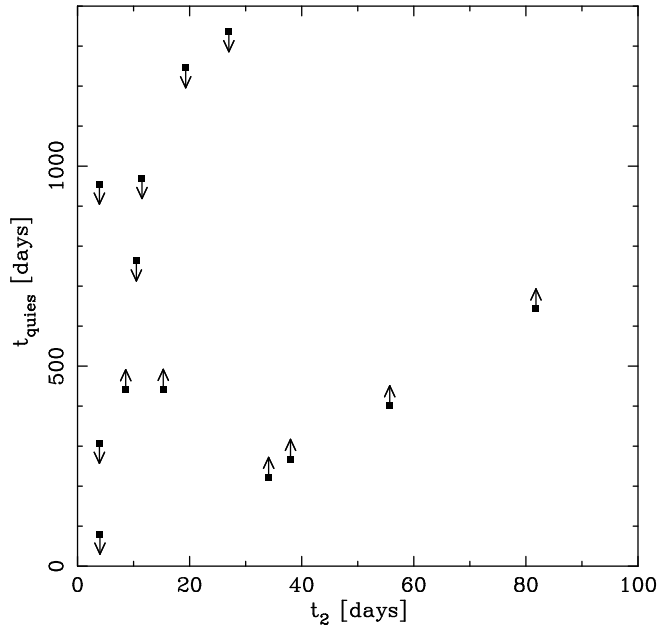}
\caption{Upper and lower limits from {\it HST} data on the time it took the novae from \citetalias{2014ApJS..213...10W} to reach quiescence.\label{fig:tquies}}
\end{figure}

To simulate the data analysis employed in \citetalias{2014ApJS..213...10W}, we must determine the availability of archival {\it HST} data based on the time of eruption of each simulated nova.  For instance, a given simulated nova will either have no archival data, only pre-eruption data, only post-eruption data, or both pre- and post-eruption data.  In the case where only post-eruption data are available we needed to determine a criterion to ensure that only suitable post-eruption data would be used, so the simulated novae had ample time to return to quiescence.  This criterion was also required to be consistent with the methodology employed in \citetalias{2014ApJS..213...10W}. The analysis undertaken in \citetalias{2014ApJS..213...10W} gave us access to post-eruption {\it HST} observations of M31 novae, which we can supplement with the additional post-eruption data analyzed by \citet{2014A&A...563L...9D} and \citet{williamsPhD}. Using these data we find that, on average, M31 novae have been confirmed to be in quiescence (or faded beyond detection) $\sim800$~days post-eruption. Therefore, we apply the simple assumption that all novae will have returned to quiescence by 800~days post-eruption. This is probably not the true picture, and one might expect the time it takes novae to reach quiescence may be correlated to a certain extent with their $t_2$ value. The upper and lower limits, which are derived from {\it HST} data, on the time it took the novae from \citetalias{2014ApJS..213...10W}, \citet{2014A&A...563L...9D} and \citet{williamsPhD} to return to quiescence are shown in Figure~\ref{fig:tquies} (note that no nova was actually observed to be detectable in eruption beyond 700~days).

\section{Modeling the RG-nova Population}\label{themodel}
Using the \citetalias{2006MNRAS.369..257D} model of the M31 nova population, coupled with our simulation of the data analysis carried out in \citetalias{2014ApJS..213...10W} we have performed Monte Carlo simulations to predict the contribution of RG-novae to the overall population of M31 novae.  A total of $1.3 \times 10^{9}$ novae were generated by the \citetalias{2006MNRAS.369..257D} model and the Monte Carlo simulation was run to generate independent random groups of 33 novae (mirroring the 33 novae contained in the ACS/WFC part of the \citetalias{2014ApJS..213...10W} catalog).  For each generated nova, the \citetalias{2006MNRAS.369..257D} model reported its position and whether it belonged to the disk or bulge nova population.  In addition, the \citetalias{2006MNRAS.369..257D} model was modified to take account of the observability (as prescribed in Section~\ref{sec:t2dist}) assigning each nova a speed class ($t_{2}$) and a peak magnitude, and to enable a given proportion of the nova parent population to be RG-novae.  As such, the RG-nova eruption probability, $\Psi^{\mathrm{RG}}$, is defined as follows

\begin{equation}
\Psi^{\mathrm{RG}}_{i} = \frac{\theta\phi^{d} f_{i}^{d} + \phi^{b}f_{i}^{b}}{\theta\sum_{j} f_{j}^{d} + \sum_{j} f_{j}^{b}},
\end{equation}

\noindent where $\phi^{d}$ and $\phi^{b}$ are the proportion of disk and bulge novae, respectively, that are RG-novae.  From this, it follows that the non-RG-nova eruption probability, $\Psi^{\prime}$, can be written as

\begin{equation}
\Psi^{\prime}_{i} = \frac{\theta\left(1-\phi^{d}\right) f_{i}^{d} + \left(1-\phi^{b}\right)f_{i}^{b}}{\theta\sum_{j} f_{j}^{d} + \sum_{j} f_{j}^{b}},
\end{equation}

\noindent and that the total nova eruption probability is given by

\begin{equation}
\Psi_{i}=\Psi^{\prime}_{i} + \Psi^{\mathrm{RG}}_{i}.
\end{equation}

If we make the assumption that the RG-nova population is drawn from, and follows, the nova population of M31 then  $\phi^{d}=\phi^{b}$.  In the case where there is no RG-nova population, then $\phi^{d}=\phi^{b}=0$ and the \citetalias{2006MNRAS.369..257D} model (Equation~\ref{DBK06model}) is recovered.

\subsection{Self-consistency}\label{self-const}
\begin{figure}
\includegraphics[width=\columnwidth]{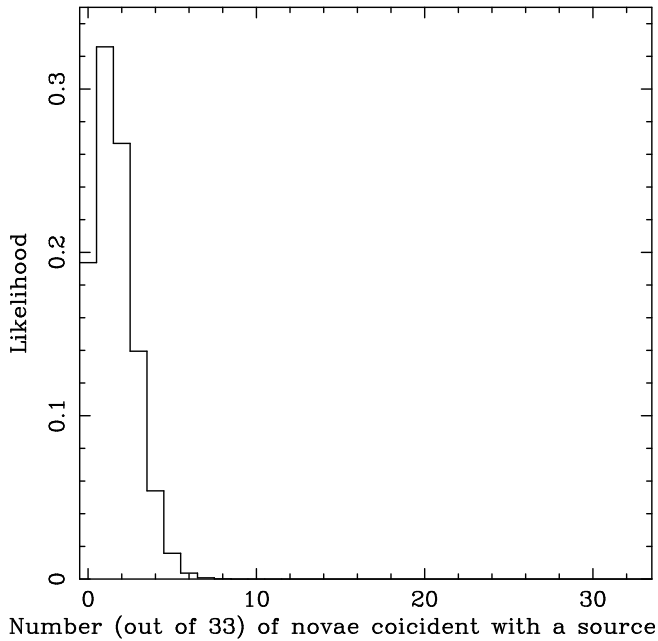}
\caption{A distribution showing the likely number of novae (out of a group of 33) that would be coincident with a resolvable stellar source in the ACS/WFC {\it HST} data if no novae were actually associated with such sources.\label{fig:norg}}
\end{figure}

As a self-consistency test, we first ran the entire Monte Carlo simulation with the RG-nova population forced to be exactly zero, i.e.\ $\phi^{d}=\phi^{b}=0$. As such, any resolved system coincident with any of these seeded novae is certain to be a chance alignment (a false positive).  As was discussed in \citetalias{2014ApJS..213...10W}, by setting a confidence limit on chance alignments at 5 percent, we would expect, on average, 5 percent of non-RG-novae would appear to have a resolved progenitor system due to chance alignments (the false positives). The results from this simulation (reproduced in Figure~\ref{fig:norg}) show that if there were no RG-novae in M31, we can say at the 78\% confidence level, our survey of 33 eruptions would expect to find no more than two false positives (or no more than 5 at the 99\% confidence level). In addition, this simulation shows that the probability of the result from \citetalias{2014ApJS..213...10W}, i.e.\ eight of 33 novae having resolved progenitors, occurring if there are no RG-novae is as small as $10^{-4}$ (excluded beyond $4\sigma$).

\subsection{Basic Model}
The basic simulation assumes the novae follow the \citetalias{2006MNRAS.369..257D} distribution and that a fixed, but unknown, proportion of the parent nova population are RG-novae, i.e.\ $\phi=\phi^{d}=\phi^{b}$.  No account for any observational biases is taken, and as such, the time of eruption, speed class, and luminosity of each generated nova are ignored by this simulation.  The RG-nova proportion, $\phi$, is varied in one percent steps between $0\le\phi\le1$.

\begin{figure}
\includegraphics[width=\columnwidth]{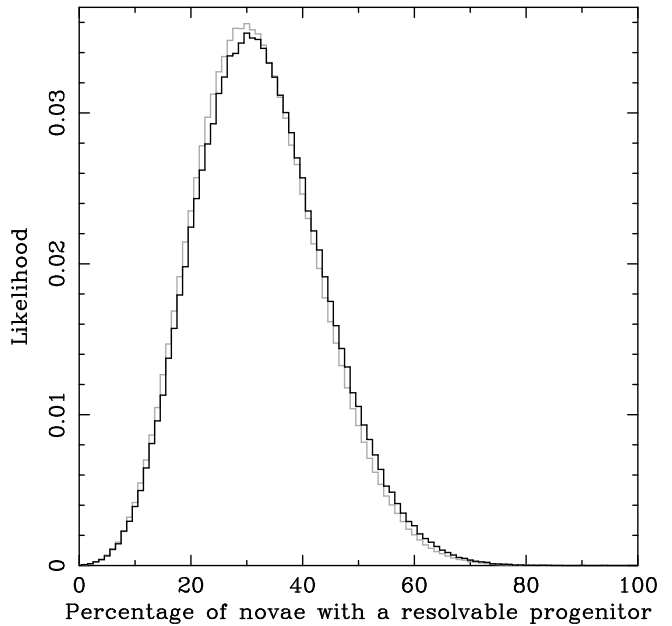}
\caption{A plot showing the distribution of likelihood probabilities over the full range of RG-nova population proportions, $0\le\phi\le1$ (resolution $\phi=0.01$).  The gray line shows the results of the basic simulation, and the black line the results of the detailed simulation that includes the effect of observational biases.\label{fig:monte}}
\end{figure}

The results of the basic simulation are plotted in Figure~\ref{fig:monte} (gray line), they indicate the most likely scenario is that the RG-nova population comprises 30 percent of the global nova population.  The confidence limits on this analysis are summarized in Table~\ref{tab:rates}.  We can constrain the RG-nova population proportion to $\phi=0.30^{+0.12}_{-0.10}$ at the $1\sigma$ equivalent level (68 percent confidence limits), and at the 99 percent confidence level we can state that the RG-nova proportion must be $>10$ percent of the global nova population.

\begin{deluxetable}{lcc}
\tablewidth{\columnwidth}
\tablecaption{Proportion of M31 nova progenitor systems expected to be resolvable in the ACS/WFC {\it HST} data from the Monte Carlo simulation described in the text.\label{tab:rates}}
\tablehead{
\colhead{Confidence} & \multicolumn{2}{c}{Proportion of M31 novae with}\\
\colhead{level} & \multicolumn{2}{c}{a resolvable progenitor, $\phi$}\\
& \colhead{Basic model\tablenotemark{1}} & \colhead{Detailed model\tablenotemark{1}}
}
\startdata
68~percent & $0.30^{+0.12}_{-0.10}$ & $0.30^{+0.13}_{-0.10}$\\
95~percent & $0.30^{+0.25}_{-0.18}$ & $0.30^{+0.26}_{-0.18}$\\
99~percent & $>0.10$ & $>0.10$
\enddata
\tablenotetext{1}{The basic model includes the \citetalias{2006MNRAS.369..257D} nova eruption model and a simulation of the data analysis undertaken in \citetalias{2014ApJS..213...10W}, whereas the detailed model also includes a simulation of the observational biases discussed in the text.}
\end{deluxetable}

\subsection{Detailed Model}\label{detailed}
The detailed simulation was run in an identical fashion to the basic simulation, but with the effect of the observational biases outlined in Section~\ref{sec:dataan} included.  Unlike the basic model, where all $1.3 \times 10^{9}$ novae are included in the simulation, the addition of the biases has a filtering effect on the simulated novae.  Here, the time of eruption, speed class, and luminosity of each generated nova is used to determine the observability of each eruption.  As such, random independent groups of 33 novae are  drawn from the novae remaining post-filtering, mirroring, as closely as possible, the conditions of the original survey.

The results of the detailed simulation are also plotted in Figure~\ref{fig:monte} (black line), and upon inspection, they are clearly remarkably similar to those from the basic simulation. This is likely because the two parameters that fundamentally govern the progenitor recovery efficiency -- the accuracy of the positional transformations and the resolved stellar density -- do not change between the two models. However, the biases discussed above do become important when we consider the spatial distribution of the likely RG-novae in Section~\ref{sec:res}. The results of the detailed model indicate that the most likely contribution of RG-novae to the overall population is 30 percent, again, the confidence limits are summarized in Table~\ref{tab:rates}.  The detailed simulation constrains the RG-nova population proportion to $\phi=0.30^{+0.13}_{-0.10}$, and at the 99 percent level the RG-nova proportion must be $>10$ percent. 	

\section{RG-Nova spatial distribution} \label{sec:res}
Throughout the analysis we have made the underlying assumption that any M31 RG-nova population belongs to the underlying parent nova population and hence follows the same spatial distribution.  Given our simulations have strongly confirmed the presence of a significant M31 RG-nova population, at the $\sim30$ percent level, we can compare its spatial distribution to that of the parent nova population.  We did this by directly comparing the spatial distribution of the likely RG-novae identified in the \citetalias{2014ApJS..213...10W} catalog with the spatial distribution of the M31 nova parent population, as computed by the \citetalias{2006MNRAS.369..257D} model, but corrected to give a population similar to the \citetalias{2014ApJS..213...10W} input catalog (correcting for all the biases discussed in Section~\ref{sec:biases}).

\begin{figure}
\begin{center}
\includegraphics[width=\columnwidth]{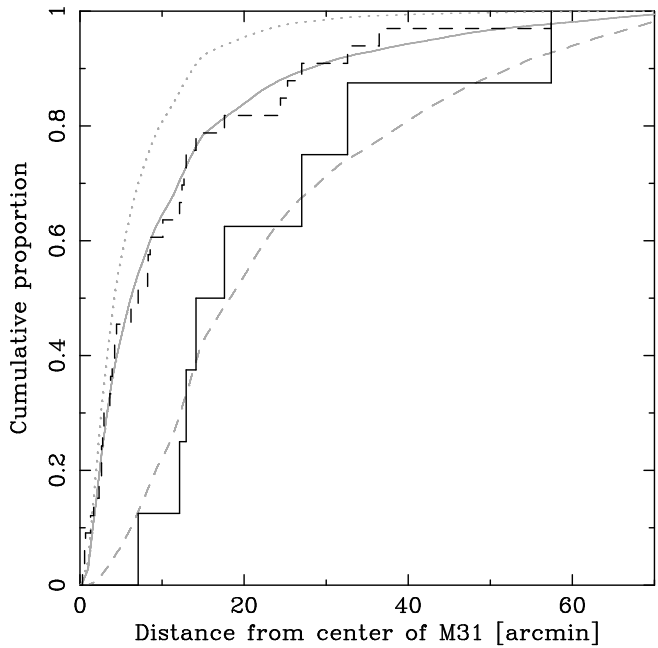}
\end{center}
\caption{The apparent distance from the center of M31 of the eight progenitor candidates identified in ACS/WFC data from LT eruption images, compared to that expected if such systems were associated with the bulge, the disk, or the entire stellar population of M31. The solid black line represents the eight RG-nova candidates used in this analysis. The dashed black line represents the distribution of all 33 novae with ACS/WFC data. The solid gray line shows the bias corrected \citetalias{2006MNRAS.369..257D} model distribution of M31 novae (i.e.\ both bulge and disk population). The dashed gray line shows the expected distribution of disk novae, with the dotted gray line representing that expected of bulge novae in the \citetalias{2014ApJS..213...10W} survey.\label{bulgedisk}}
\end{figure}

In Figure~\ref{bulgedisk} we present a plot of the cumulative spatial distribution (based on apparent distance from the center of M31) of the eight RG-novae found in ACS data from the LT eruption image in \citetalias{2014ApJS..213...10W} (the positions of these systems are also plotted in Figure~\ref{spatial}).  In Figure~\ref{bulgedisk}, the RG-nova spatial distribution can first be compared directly to that of the underlying nova population, from the bias corrected \citetalias{2006MNRAS.369..257D} model, the solid gray line represents a ratio of disk and bulge nova eruption rates of $\theta=0.18$, as was assumed throughout the simulations.  An AD-test between the spatial distributions of the RG-novae and the general M31 parent nova population suggests the probability that these two distributions are drawn from the same population is only 0.006. Conversely, Figure~\ref{bulgedisk} shows the the cumulative distribution of all 33 novae used in the analysis is described well by the model for the overall nova distribution (AD-test P=0.91). This would of course be expected, as Section~\ref{sec:res} uses the observational data for M31 novae occurring over a number of years to match the nova model with the distribution we would actually expect to see in the input catalog, indeed the nova sample used to determine this would have included these 33 systems.

It should also be noted that any chance alignments in the survey published in \citetalias{2014ApJS..213...10W} (false positives), are likely to shift the distribution of RG-novae back towards the overall M31 nova distribution (solid gray line in Figure~\ref{bulgedisk}). One may imagine a nova occurring near the center of M31 may have a progenitor unresolvable in the data, that may have been resolvable if it had resided in the lower surface-brightness outer regions of the galaxy. This effect is likely to be relatively small, as the limiting magnitude is still quite faint until very close to the center of M31. Another (probably small) effect may be the stellar density, where progenitor recovery may become less efficient in areas of higher stellar density.

The small sample size provided by the \citetalias{2014ApJS..213...10W} catalog (8 RG-novae with ACS/WFC data analyzed here) effectively prohibits detailed exploration of separate bulge and disk RG-nova populations (i.e.\ $\phi^{d}\neq\phi^{b}$).  However, we can investigate further a number of special cases, that where the bulge contribution to the RG-nova rate is exactly zero, and that where the disk contribution to the RG-nova rate is zero, e.g. $\phi^{d}\neq\phi^{b}=0$ and $\phi^{b}\neq\phi^{d}=0$.  Figure~\ref{bulgedisk} compares the RG-nova spatial distribution from \citetalias{2014ApJS..213...10W} with the spatial distribution of the M31 disk nova population (gray dashed line) and the bulge nova population (dotted line).  An AD-test between the bulge and RG-nova spatial distributions gives the probability that these two distributions are drawn from the same population as $10^{-6}$, effectively ruling this possibility out. The same statistic for the disk and RG-nova distributions computes a probability of 0.96. When we compare the spatial distribution of all eleven RG-nova candidates identified in \citetalias{2014ApJS..213...10W} to the model distribution shown in Figure~\ref{bulgedisk}, the AD-test still indicates that the RG-novae are unlikely to follow the overall nova distribution ($P=0.02$). Simply comparing the 8 RG-novae to the distribution of all 33 novae, an AD-test yields a probability of 0.01 the two have the same parent populations.

\begin{figure}
\includegraphics[width=\columnwidth]{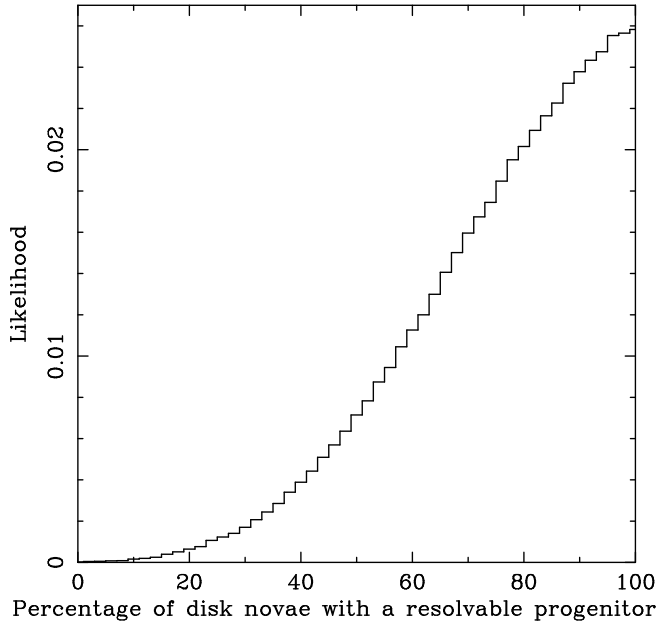}
\caption{A plot showing the distribution of likelihood probabilities over the full range of disk RG-nova population proportions, $0\le\phi^{d}\le1$ (resolution $\phi^{d}=0.02$) assuming that there is no contribution to the RG-nova population from the bulge novae ($\phi^{b}=0$).\label{fig:diskrg}}
\end{figure}

As such, it seems clear that any RG-nova in M31 follows the distribution of the disk novae more closely than that of the overall nova population.  If we therefore make the assumption that the bulge RG-nova proportion is exactly zero, $\phi^{b}=0$, then we can modify our simulations to put some constraints on the disk RG-nova proportion, $\phi^{d}$.  The detailed simulation (see Section~\ref{detailed}) was re-run with the above constraints, and the results are plotted in Figure~\ref{fig:diskrg}.  Our results indicate, the somewhat surprising picture, that the most likely conclusion is the disk nova population is dominated by RG-novae.  In this scenario, with the RG-novae just occurring in the disk, we can contain the disk RG-nova proportion at $\phi^{d}>0.68$, $>0.40$, and $>0.26$ at the 68, 95, and 99 percent confidence limits, respectively. Given that (again taking $\theta=0.18$) the \citetalias{2006MNRAS.369..257D} model predicts the overall M31 bulge/disk nova ratio to be 4:3, this would imply a global RG-nova proportion of $>0.29$ at the 68 percent confidence level. We stress however that these numbers would only be valid if no RG-novae reside in the M31 bulge.

\section{Discussion}
The full statistical analysis of the results of the survey for quiescent novae in M31 (\citetalias{2014ApJS..213...10W}) suggests that $0.30^{+0.13}_{-0.10}$ of all M31 novae have progenitors that would be resolvable in ACS/WFC {\it HST} data, a group of systems expected to be dominated by RG-novae. This is an important result for the study of nova populations, as it shows a significant proportion of all M31 novae likely contain evolved secondaries. It is also an important result when considering the RG-nova SN Ia progenitor channel. Previously only about three percent of the approximately 400 novae in our Galaxy were confirmed or suspected RG-novae. The significantly higher M31 RG-nova rate our results imply means that such novae could potentially make a more significant contribution to the SN Ia rate than previously believed.

Recent work by \citet{2014ApJ...788..164P} attempted to identify possible RNe amongst the Galactic CN population. They suggest that 17 of the 75 CNe (23~percent) where the likely nature of the secondary was able to be determined were probably RG-novae. However, this number excludes the numerous eruptions produced by recurrent RG-novae such as RS~Oph, V745~Scorpii and T~Coronae Borealis. Of the 40 known Galactic RN eruptions, 17 can be attributed to RG-novae, 15 to SG-novae and eight to MS-novae (see \citealp{2010ApJS..187..275S} and \citealp{2012ApJ...746...61D}). It should be noted that with only ten confirmed Galactic RN systems, these proportions can be greatly influenced by a single system with a short recurrence time. For example, between them, U~Scorpii (SG-nova), RS~Oph (RG-nova) and T~Pyxidis (MS-nova) contribute 23 of the 40 aforementioned eruptions.

From the work on Galactic novae published by \citet{2014ApJ...788..164P}, we can calculate that of the 115 Galactic nova eruptions (CNe and RNe) where the evolutionary state of the secondary is known 34 (30~percent) occur in RG-nova systems. Initially this seems to agree very well with our findings in M31. However, one of the main methods of determining the nature of the secondary by \citet{2014ApJ...788..164P} are the infrared colors. Therefore as MS-novae will be significantly less luminous in the infrared, this will likely lead to a situation where at a given distance, MS-novae are harder to classify due to not being resolved in infrared imaging and thus more likely to be excluded from the above statistic. In addition, if the spatial distribution found in our sample of M31 novae (see Section~\ref{sec:res}) is replicated in our Galaxy, one may expect to find an even higher rate, due to the nearby (and thus disk) novae being brighter and therefore easier to discover, particularly when looking at historical data. Given the uncertainty in the M31 RG-nova rate (0.12 to 0.56 at the 95 percent confidence level) and the lack of an estimate of the true RG-nova rate in our Galaxy, there is no reason to believe the two are incompatible. This does illustrate the need for both a significantly larger sample size of M31 nova progenitors and deep multi-color infrared imaging of old Galactic novae.

The relatively small sample size of M31 RG-novae effectively prevents a detailed exploration of the proportion of systems associated with the disk and bulge of M31. However, we can say with a high degree of confidence that a higher proportion of M31 disk novae are RG-novae than those found in the bulge. Indeed our result is consistent with RG-novae only residing in the disk. This would be in contrast to the overall M31 nova population, where the majority appear to be associated with the bulge (\citealp{2001ApJ...563..749S}; \citetalias{2006MNRAS.369..257D}), and strongly hints that RG-novae arise from a separate (younger) stellar population.  The results of the statistical analysis do not preclude all disk novae being RG-novae; although we should stress that such an extreme situation seems very unlikely. Indeed, if for example we consider M31N~2009-08b, this is about $36'$ from the M31 center and is highly likely to reside in the disk, we see the progenitor search in \citetalias{2014ApJS..213...10W} effectively rules out this being a RG-nova. The fact that our results indicate that RG-novae may follow the M31 disk light is an important result with respect to RG-novae as potential progenitors of SNe~Ia, as, if confirmed in other galaxies, it implies SNe~Ia arising from RG-nova systems would be much more likely to occur in younger stellar populations, rather than old stellar populations such as early-type galaxies.  This outcome can clearly be tested in Local Group galaxies dominated by younger stellar populations, such as M33 and the Magellanic Clouds.

The presence of narrow blueshifted time-varying {Na~{\sc i}} doublet absorption features in the spectra of SN Ia has been suggested as evidence of a circumstellar medium, and therefore of a red-giant wind or prior RN eruptions \citep{2007Sci...317..924P,2009ApJ...702.1157S}, although it has also been suggested that double-degenerate progenitors may be capable of producing such features \citep{2013MNRAS.431.1541S}. Observations of these {Na~{\sc i}} doublet absorption features in SN Ia by \citet[][see also \citealp{2011Sci...333..856S}]{2013MNRAS.436..222M} suggests that at least 20 percent of SN Ia have a contribution from a circumstellar medium. Notably, none of the SN Ia with such absorption features in their sample were found in early-type galaxies. All SNe Ia-CSM appear to reside in late-type spirals or dwarf irregulars \citep{2013ApJS..207....3S}.

Our results suggest that any observational signatures found in SNe Ia originating from RG-nova systems would be expected to be more likely to be found in disk galaxies and may be largely absent from SN Ia arising from old stellar populations.

\subsection{Parameter sensitivity}\label{paramsens}
\begin{figure}
\includegraphics[width=\columnwidth]{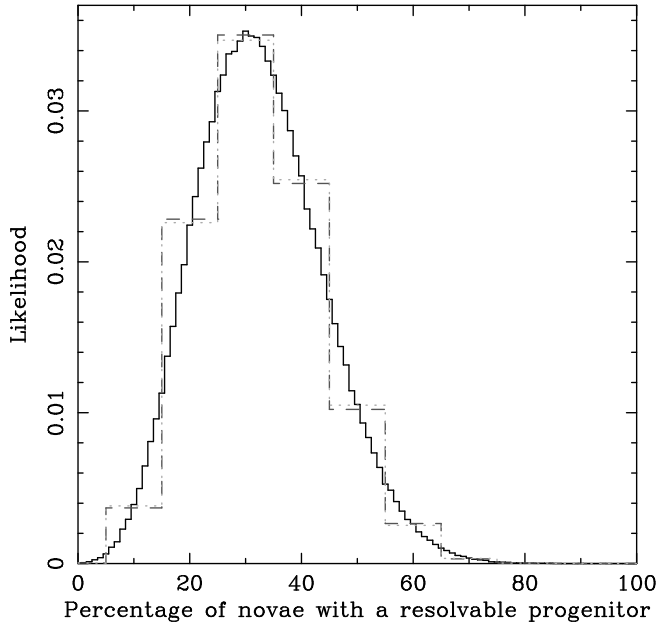}
\caption{A plot showing the distribution of likelihood probabilities over the full range of RG-nova population proportions, $0\le\phi\le1$.  The black line (resolution $\phi=0.01$) shows the results of the detailed simulations for $\theta=0.18$, the dark gray dashed line and light gray dotted line (both resolution $\phi=0.10$) show the distribution for $\theta=0.42$ and $\theta=0.08$, respectively.\label{fig:diskprop}}
\end{figure}

As discussed briefly in Section~\ref{sec:spa}, the \citetalias{2006MNRAS.369..257D} M31 nova eruption model relies upon a single parameter, $\theta$, the ratio of disk and bulge nova eruption rates.  \citetalias{2006MNRAS.369..257D} constrain this parameter to be $\theta=0.18^{+0.24}_{-0.10}$.  However, for our analysis we assumed a value of $\theta=0.18$ throughout.  To investigate the sensitivity of our analysis to the choice of $\theta$ we re-ran the detailed simulation (albeit, at a lower resolution) for values of $\theta=0.08$ and $\theta=0.42$, the range suggested by the \citetalias{2006MNRAS.369..257D} analysis.  The results of these additional simulations are shown in Figure~\ref{fig:diskprop}, where the $\theta=0.08$ (dashed line) and $\theta=0.42$ (dotted line) results are compared directly to the results of the detailed simulation (solid line).  The results from all three simulations are remarkably similar, and confirm that our general results are not sensitive to the choice of $\theta$, within the range constrained by \citetalias{2006MNRAS.369..257D}.

\subsection{Object nature}\label{nature}
\begin{figure*}
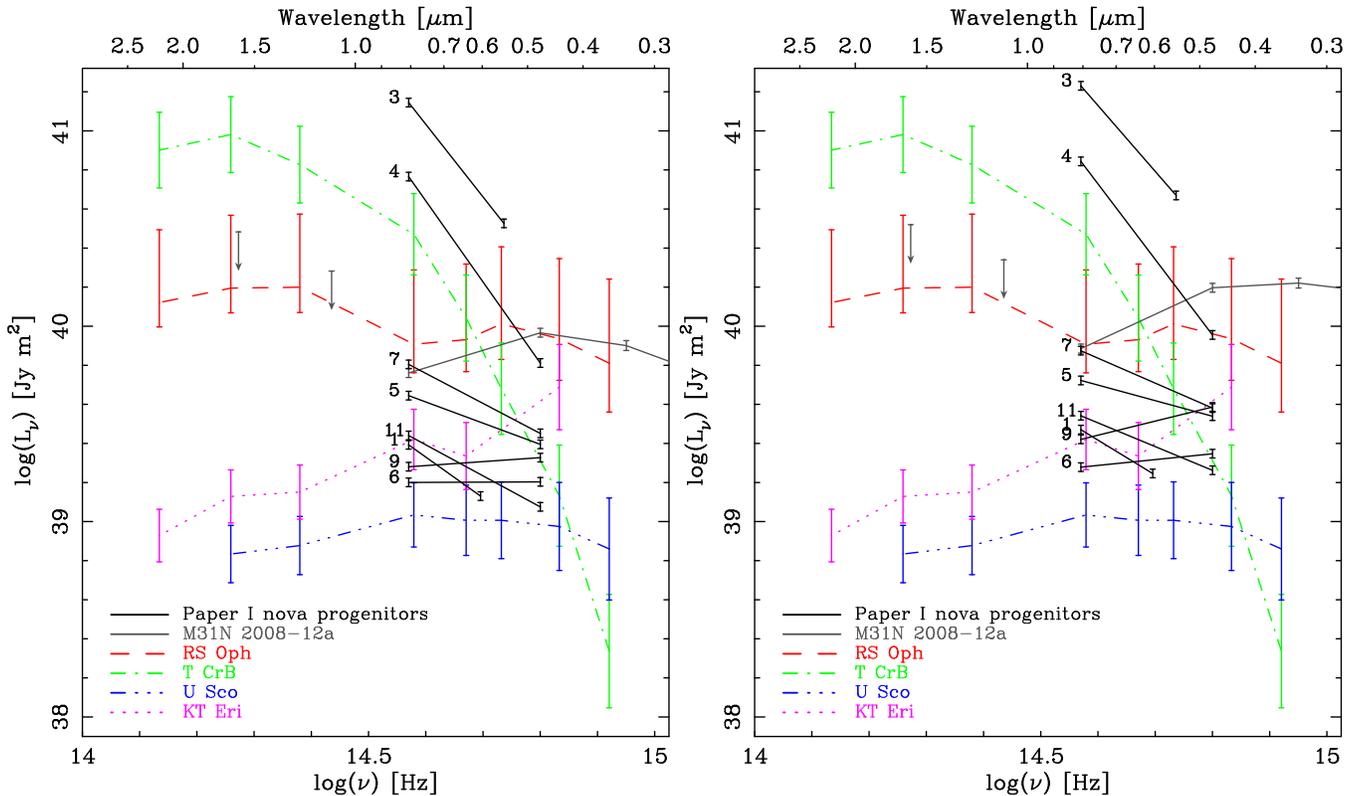

\begin{center}
\includegraphics[width=0.49\textwidth]{sed_lowext}
\includegraphics[width=0.49\textwidth]{sed_highext}
\end{center}
\caption{Distance and extinction corrected SEDs showing the progenitor systems of the quiescent Galactic RNe T~CrB, RS~Oph, KT~Eri, and U~Sco (the green, red, magenta, and blue data, respectively, see also the figure key). The gray data indicates the quiescent M31 1~year period recurrent M31N~2008-12a \citep[see,][]{2014A&A...563L...9D}, and the black data indicate the \citetalias{2014ApJS..213...10W} systems (1)~M31N~2007-02b, (3)~2007-11b, (4)~2007-11d, (5)~2007-11e, (6)~2007-12a, (7)~2007-12b, (9)~2009-11d, and (11)~2010-09b.  Units are chosen for comparison with similar plots in \citet[see their Figure~71]{2010ApJS..187..275S} and \citet[see their Figures~4 and 11, respectively]{2014A&A...563L...9D,2015A&A...580A..45D}.  The left-hand plot is the low extinction scenario, where only the line-of-sight (Galactic) extinction towards M31 is considered \citep[$E_{B-V}=0.1$;][]{1992ApJS...79...77S}.  The right-hand plot considers an additional extinction internal to M31, as shown for each \citetalias{2014ApJS..213...10W} nova in Table~\ref{tab:reddening}.\label{fig:sed}}
\end{figure*}

As was discussed in Section~\ref{sec:two}, throughout this paper we have worked under the assumption that all eleven progenitor systems identified in \citetalias{2014ApJS..213...10W} contained red giant secondaries; RG-novae.  Figure~\ref{fig:sed} presents the NUV, optical, and IR spectral energy distributions (SEDs) of three quiescent Galactic RNe, and one Galactic RN candidate, compared to the quiescent SED of the one-year recurrence period M31N~2008-12a, and eight of the eleven \citetalias{2014ApJS..213...10W} novae (those for which quiescent colors are available; see Tables~4 and 5 of \citetalias{2014ApJS..213...10W}).  For the Galactic RNe we have selected: RS~Oph (red line), a RG-nova with a luminous accretion disk; T~CrB (green), a RG-nova with no obvious disk; and U~Sco (blue), a SG-nova with a luminous accretion disk, one of the most luminous accretion dominated systems.  We have also included the SED of the Galactic RN candidate, and proposed RG-nova  KT~Eridani \citep[magenta line; see][]{2012A&A...537A..34J}. The quiescent photometry for the Galactic RNe is taken from \citet[][see their Table~30]{2010ApJS..187..275S}, the quiescent photometry for KT~Eri is from \citet{2012A&A...537A..34J}, and the distances and reddening are from \citet[][see their Table~2 and references within]{2012ApJ...746...61D}.  We have assumed a distance towards RS~Oph of $1.4^{+0.6}_{-0.2}$\,kpc (\citealp{2008ASPC..401...52B}; see also \citealp{1987rorn.conf..241B}).  For the M31 novae we have assumed a distance to M31 of $770\pm19$\,kpc \citep{1990ApJ...365..186F}, a line-of-sight Galactic (external) reddening towards M31 of $E_{B-V}=0.1$ \citep{1992ApJS...79...77S}, and an upper limit on the reddening internal to M31 as given for the position of each object by the M31 reddening map of \citet[][see Table~\ref{tab:reddening}]{2009A&A...507..283M}.

\begin{deluxetable}{llr}
\tablewidth{\columnwidth}
\tablecaption{Estimate of maximum internal M31 reddening towards each RG-nova.\label{tab:reddening}}
\tablehead{
\colhead{Number\tablenotemark{a}} & \colhead{Nova} & \colhead{$E_{\mathrm{B-V}}^{\mathrm{internal}}$}
}
\startdata
1 & M31N~2007-02b & 0.10 \\
2 & M31N~2007-10a & 0.16 \\
3 & M31N~2007-11b & 0.11 \\
4 & M31N~2007-11d & 0.10 \\
5 & M31N~2007-11e & 0.10\tablenotemark{b} \\
6 & M31N~2007-12a & 0.10 \\
7 & M31N~2007-12b & 0.09 \\
8 & M31N~2009-08a & 0.08 \\
9 & M31N~2009-11d & 0.18 \\
10 & M31N~2010-01a & 0.08 \\
11 & M31N~2010-09b & 0.13 
\enddata
\tablenotetext{a}{Numbers simply refer to the SEDs plotted in Figure~\ref{fig:sed}.}
\tablenotetext{b}{No reddening data are available at the position of M31N~2007-11e in the \citet{2009A&A...507..283M} reddening map.  Here we take the mean reddening in the $1^{\prime}\times1^{\prime}$ region around the nova.}
\end{deluxetable}

The quiescent SEDs of M31N~2007-02b, 2007-11b, 2007-11d, 2007-11e, 2007-12b, and 2010-09b (see SED \#1, 3, 4, 5, 7, and 11, respectively, in Figure~\ref{fig:sed}; note M31N~2007-11b is not one of the RG-novae included in the statistical analysis) are all consistent with the expected luminosity and slope of a system dominated by a red giant star (at the distance of M31), i.e., they are consistent with the quiescent SED of T~CrB.  The quiescent SEDs of M31N~2007-12a and 2009-11d (\#6 and 9, respectively) however, show evidence of the presence of a luminous accretion disk, and these SEDs are more consistent with that of KT~Eri, which contains a low luminosity red giant secondary and a bright accretion disk.  Unfortunately, the quiescent photometry for M31N~2007-10a (\#2) consists of only a single filter, and as such, is not plotted in Figure~\ref{fig:sed}.  However the luminosity of this system is consistent with either a red giant secondary or an accretion disk more luminous than that in KT~Eri.

A significant proportion of the novae in the survey erupted before the region had been covered by the PHAT survey. When calculating the RG-nova rate of M31, it is also important that all eight of the novae were in quiescence if post-eruption data were used. Of the eight, M31N~2007-11d, 2007-11e, 2007-12a, 2007-12b and 2009-11d all rely on post-eruption data. The progenitor of M31N~2007-12b was already recovered by \citet{2009ApJ...705.1056B} in pre-eruption data. The red ($V-I$) colors of the progenitors of M31N~2007-11d and 2007-11e (and indeed all but M31N~2007-12a and 2009-11d) strongly imply that we are not seeing the tail of the nova eruptions. At very late times in an eruption the optical luminosity of the eruption itself is dominated by line emission, even in broad filters, with [O~{\sc iii}] (5007\,\AA) particularly strong in green filters. This effect can be seen in the late-time light curves presented in \citetalias{2014ApJS..213...10W}. M31N~2009-11d ($t_2=10$~days; \citealp{2011ApJ...734...12S}), for which the {\it HST} data were two years after the eruption, is observed to be almost the same F475W magnitude on 2012 Jul 21 as on 2011 Dec 22 \citep{williamsPhD}. This only leaves M31N~2007-12a ($t_2=25$~days; \citealp{2011ApJ...727...50S}), for which the {\it HST} data were taken three years after the eruption.

From this analysis, we are confident that the majority, if not all, of these systems are consistent with quiescent RG-novae. However, the possibility remains that M31N~2009-11d may be a luminous accretion disk and M31N~2007-12a may be either a luminous accretion disk or the tail of the eruption. Although these systems are still likely RG-novae, we now consider our results if they are not. If we assume we are instead seeing luminous accretion disks in non-RG systems, this changes the final result from $30^{+13}_{-10}$ to $22^{+12}_{-9}$ percent. Assuming we are seeing the tail of the M31N~2007-12a eruption (thereby removing the system from the analysis), but all other systems are either RG-novae or chance alignments, changes the final result to $27^{+13}_{-10}$ percent. Note that it is possible a bias towards non-RG-novae may be introduced by the use of post-eruption data. For example, a non-detection can clearly be interpreted as the system not being a RG-nova, however if a coincident source is detected, in some cases it may not be clear if it is the quiescent system or the tail of an eruption, and thus excluded from the survey.

\section{Conclusions}
In this paper we have conducted a statistical analysis of the M31 nova progenitor survey published in \citetalias{2014ApJS..213...10W}, and considered the possible biases in the catalog. Here we summarize the main conclusions of our analysis. 

\begin{itemize}
\item Based on the distribution of decline times ($t_{2}$) the nova population of M31 is not similar to the observed Galactic nova population.  However, the disk nova population of M31 (based on all novae erupting $>15^{\prime}$ from the center of M31) may be similar to the observed Galactic population.
\item A statistical analysis of the progenitor survey conducted in \citetalias{2014ApJS..213...10W} predicts that $30^{+13}_{-10}$ percent of all M31 nova eruptions can be attributed to RG-nova systems. This is the first determination of the RG-nova rate of an entire galaxy. The Galactic RG-nova proportion has traditionally stood at $\sim3$~percent, although recent results have indicated it may be much higher.
\item This elevated RG-nova rate means that such systems may be responsible for a higher proportion of SN~Ia explosions than previously thought.
\item The RG-novae appear to be more strongly associated with the M31 disk than the bulge. Indeed, our results are consistent with such systems being entirely associated with the disk population.
\item We therefore predict that any SNe~Ia originating from RG-nova systems are significantly less likely to be associated with old stellar populations, and will mainly occur in younger populations, such as galactic disks.
\end{itemize}

\acknowledgements The Liverpool Telescope is operated on the island of La Palma by Liverpool John Moores University in the Spanish Observatorio del Roque de los Muchachos of the Instituto de Astrofisica de Canarias with financial support from the UK Science and Technology Facilities Council (STFC).  Some of the data presented in this paper were obtained from the Multimission Archive at the Space Telescope Science Institute (MAST).  STScI is operated by the Association of Universities for Research in Astronomy, Inc., under NASA contract NAS5-26555.  Support for MAST for non-{\it HST} data is provided by the NASA Office of Space Science via grant NNX09AF08G and by other grants and contracts. S.C.W.\ acknowledges support from an STFC Ph.D.\ studentship and would like to thank D.\ Fraquelli for advice on using the Starview program, which is hosted by STScI.  M.J.D.\ would like to thank M.\ Henze for helpful discussions about statistics. A.W.S.\ acknowledges support from NSF grant AST1009566. The authors would also like to thank A.\ Newsam and T.\ O'Brien for invaluable discussion and advice.  Finally, the authors would also like to offer their thanks to an anonymous referee whose comments helped improve the original manuscript.

{\it Facilities: \facility{HST}, \facility{HET}}, \facility{Liverpool:2m}.

\bibliographystyle{apj}
\bibliography{refs}

\begin{thebibliography}{}
\expandafter\ifx\csname natexlab\endcsname\relax\def\natexlab#1{#1}\fi

\bibitem[{{Anderson} \& {Darling}(1952)}]{anderson1952}
{Anderson}, T.~W., \& {Darling}, D.~A. 1952, Ann. Math. Statist., 23, 193

\bibitem[{{Barry} {et~al.}(2008){Barry}, {Mukai}, {Sokoloski}, {Danchi},
  {Hachisu}, {Evans}, {Gehrz}, \& {Mikolajewska}}]{2008ASPC..401...52B}
{Barry}, R.~K., {Mukai}, K., {Sokoloski}, J.~L., {et~al.} 2008, in Astronomical
  Society of the Pacific Conference Series, Vol. 401, RS Ophiuchi (2006) and
  the Recurrent Nova Phenomenon, ed. A.~{Evans}, M.~F. {Bode}, T.~J. {O'Brien},
  \& M.~J. {Darnley} (San Francisco: Astronomical Society of the Pacific), 52

\bibitem[{{Bode}(1987)}]{1987rorn.conf..241B}
{Bode}, M.~F. 1987, in RS Ophiuchi (1985) and the Recurrent Nova Phenomenon,
  ed. M.~F. {Bode} (Utrecht: VNU Science Press), 241

\bibitem[{{Bode}(2010)}]{2010AN....331..160B}
{Bode}, M.~F. 2010, Astronomische Nachrichten, 331, 160

\bibitem[{{Bode} {et~al.}(2009){Bode}, {Darnley}, {Shafter}, {Page},
  {Smirnova}, {Anupama}, \& {Hilton}}]{2009ApJ...705.1056B}
{Bode}, M.~F., {Darnley}, M.~J., {Shafter}, A.~W., {et~al.} 2009, \apj, 705,
  1056

\bibitem[{{Bode} \& {Evans}(2008)}]{2008clno.book.....B}
{Bode}, M.~F., \& {Evans}, A., eds. 2008, Cambridge Astrophysics Series,
  Vol.~43, {Classical Novae, 2nd Edition} (Cambridge: Cambridge University
  Press)

\bibitem[{{Bode} {et~al.}(2006){Bode}, {O'Brien}, {Osborne}, {Page},
  {Senziani}, {Skinner}, {Starrfield}, {Ness}, {Drake}, {Schwarz}, {Beardmore},
  {Darnley}, {Eyres}, {Evans}, {Gehrels}, {Goad}, {Jean}, {Krautter}, \&
  {Novara}}]{2006ApJ...652..629B}
{Bode}, M.~F., {O'Brien}, T.~J., {Osborne}, J.~P., {et~al.} 2006, \apj, 652,
  629

\bibitem[{{Bode} {et~al.}(2008){Bode}, {Osborne}, {Page}, {Beardmore},
  {O'Brien}, {Ness}, {Starrfield}, {Skinner}, {Darnley}, {Drake}, {Evans},
  {Eyres}, {Krautter}, \& {Schwarz}}]{2008ASPC..401..269B}
{Bode}, M.~F., {Osborne}, J.~P., {Page}, K.~L., {et~al.} 2008, in Astronomical
  Society of the Pacific Conference Series, Vol. 401, RS Ophiuchi (2006) and
  the Recurrent Nova Phenomenon, ed. A.~{Evans}, M.~F. {Bode}, T.~J. {O'Brien},
  \& M.~J. {Darnley} (San Francisco: Astronomical Society of the Pacific), 269

\bibitem[{{Cao} {et~al.}(2012){Cao}, {Kasliwal}, {Neill}, {Kulkarni}, {Lou},
  {Ben-Ami}, {Bloom}, {Cenko}, {Law}, {Nugent}, {Ofek}, {Poznanski}, \&
  {Quimby}}]{2012ApJ...752..133C}
{Cao}, Y., {Kasliwal}, M.~M., {Neill}, J.~D., {et~al.} 2012, \apj, 752, 133

\bibitem[{{Ciardullo} {et~al.}(1987){Ciardullo}, {Ford}, {Neill}, {Jacoby}, \&
  {Shafter}}]{1987ApJ...318..520C}
{Ciardullo}, R., {Ford}, H.~C., {Neill}, J.~D., {Jacoby}, G.~H., \& {Shafter},
  A.~W. 1987, \apj, 318, 520

\bibitem[{{Dalcanton} {et~al.}(2012){Dalcanton}, {Williams}, {Lang}, {Lauer},
  {Kalirai}, {Seth}, {Dolphin}, {Rosenfield}, {Weisz}, {Bell}, {Bianchi},
  {Boyer}, {Caldwell}, {Dong}, {Dorman}, {Gilbert}, {Girardi}, {Gogarten},
  {Gordon}, {Guhathakurta}, {Hodge}, {Holtzman}, {Johnson}, {Larsen}, {Lewis},
  {Melbourne}, {Olsen}, {Rix}, {Rosema}, {Saha}, {Sarajedini}, {Skillman}, \&
  {Stanek}}]{2012ApJS..200...18D}
{Dalcanton}, J.~J., {Williams}, B.~F., {Lang}, D., {et~al.} 2012, \apjs, 200,
  18

\bibitem[{{Darnley}(2005)}]{2005PhDT.........2D}
{Darnley}, M.~J. 2005, PhD thesis, Astrophysics Research Institute, Liverpool
  John Moores, University, Twelve Quays House, Egerton Wharf, Birkenhead, CH41
  1LD, UK

\bibitem[{{Darnley} {et~al.}(2012){Darnley}, {Ribeiro}, {Bode}, {Hounsell}, \&
  {Williams}}]{2012ApJ...746...61D}
{Darnley}, M.~J., {Ribeiro}, V.~A.~R.~M., {Bode}, M.~F., {Hounsell}, R.~A., \&
  {Williams}, R.~P. 2012, \apj, 746, 61

\bibitem[{{Darnley} {et~al.}(2014){Darnley}, {Williams}, {Bode}, {Henze},
  {Ness}, {Shafter}, {Hornoch}, \& {Votruba}}]{2014A&A...563L...9D}
{Darnley}, M.~J., {Williams}, S.~C., {Bode}, M.~F., {et~al.} 2014, \aap, 563,
  L9

\bibitem[{{Darnley} {et~al.}(2004){Darnley}, {Bode}, {Kerins}, {Newsam}, {An},
  {Baillon}, {Novati}, {Carr}, {Cr{\'e}z{\'e}}, {Evans}, {Giraud-H{\'e}raud},
  {Gould}, {Hewett}, {Jetzer}, {Kaplan}, {Paulin-Henriksson}, {Smartt},
  {Stalin}, \& {Tsapras}}]{2004MNRAS.353..571D}
{Darnley}, M.~J., {Bode}, M.~F., {Kerins}, E., {et~al.} 2004, \mnras, 353, 571

\bibitem[{{Darnley} {et~al.}(2006){Darnley}, {Bode}, {Kerins}, {Newsam}, {An},
  {Baillon}, {Belokurov}, {Calchi Novati}, {Carr}, {Cr{\'e}z{\'e}}, {Evans},
  {Giraud-H{\'e}raud}, {Gould}, {Hewett}, {Jetzer}, {Kaplan},
  {Paulin-Henriksson}, {Smartt}, {Tsapras}, \& {Weston}}]{2006MNRAS.369..257D}
---. 2006, \mnras, 369, 257 (DBK06)

\bibitem[{{Darnley} {et~al.}(2015){Darnley}, {Henze}, {Steele}, {Bode},
  {Ribeiro}, {Rodr{\'{\i}}guez-Gil}, {Shafter}, {Williams}, {Baer}, {Hachisu},
  {Hernanz}, {Hornoch}, {Hounsell}, {Kato}, {Kiyota}, {Ku{\v c}{\'a}kov{\'a}},
  {Maehara}, {Ness}, {Piascik}, {Sala}, {Skillen}, {Smith}, \&
  {Wolf}}]{2015A&A...580A..45D}
{Darnley}, M.~J., {Henze}, M., {Steele}, I.~A., {et~al.} 2015, \aap, 580, A45

\bibitem[{{della Valle} {et~al.}(1992){della Valle}, {Bianchini}, {Livio}, \&
  {Orio}}]{1992A&A...266..232D}
{della Valle}, M., {Bianchini}, A., {Livio}, M., \& {Orio}, M. 1992, \aap, 266,
  232

\bibitem[{{Dilday} {et~al.}(2012){Dilday}, {Howell}, {Cenko}, {Silverman},
  {Nugent}, {Sullivan}, {Ben-Ami}, {Bildsten}, {Bolte}, {Endl}, {Filippenko},
  {Gnat}, {Horesh}, {Hsiao}, {Kasliwal}, {Kirkman}, {Maguire}, {Marcy},
  {Moore}, {Pan}, {Parrent}, {Podsiadlowski}, {Quimby}, {Sternberg}, {Suzuki},
  {Tytler}, {Xu}, {Bloom}, {Gal-Yam}, {Hook}, {Kulkarni}, {Law}, {Ofek},
  {Polishook}, \& {Poznanski}}]{2012Sci...337..942D}
{Dilday}, B., {Howell}, D.~A., {Cenko}, S.~B., {et~al.} 2012, Science, 337, 942

\bibitem[{{Dolphin}(2000)}]{2000PASP..112.1383D}
{Dolphin}, A.~E. 2000, \pasp, 112, 1383

\bibitem[{{Duerbeck}(1990)}]{1990LNP...369...34D}
{Duerbeck}, H.~W. 1990, in Lecture Notes in Physics, Vol. 369, IAU Colloq. 122:
  Physics of Classical Novae, ed. A.~{Cassatella} \& R.~{Viotti} (Berlin:
  Springer Verlag), 34

\bibitem[{{Freedman} \& {Madore}(1990)}]{1990ApJ...365..186F}
{Freedman}, W.~L., \& {Madore}, B.~F. 1990, \apj, 365, 186

\bibitem[{{Gutierrez} {et~al.}(1996){Gutierrez}, {Garcia-Berro}, {Iben},
  {Isern}, {Labay}, \& {Canal}}]{1996ApJ...459..701G}
{Gutierrez}, J., {Garcia-Berro}, E., {Iben}, Jr., I., {et~al.} 1996, \apj, 459,
  701

\bibitem[{{Henze} {et~al.}(2015{\natexlab{a}}){Henze}, {Darnley}, {Kabashima},
  {Nishiyama}, {Itagaki}, \& {Gao}}]{2015A&A...582L...8H}
{Henze}, M., {Darnley}, M.~J., {Kabashima}, F., {et~al.} 2015{\natexlab{a}},
  \aap, 582, L8

\bibitem[{{Henze} {et~al.}(2014{\natexlab{a}}){Henze}, {Ness}, {Darnley},
  {Bode}, {Williams}, {Shafter}, {Kato}, \& {Hachisu}}]{2014A&A...563L...8H}
{Henze}, M., {Ness}, J.-U., {Darnley}, M.~J., {et~al.} 2014{\natexlab{a}},
  \aap, 563, L8

\bibitem[{{Henze} {et~al.}(2014{\natexlab{b}}){Henze}, {Pietsch}, {Haberl},
  {Della Valle}, {Sala}, {Hatzidimitriou}, {Hofmann}, {Hernanz}, {Hartmann}, \&
  {Greiner}}]{2014A&A...563A...2H}
{Henze}, M., {Pietsch}, W., {Haberl}, F., {et~al.} 2014{\natexlab{b}}, \aap,
  563, A2

\bibitem[{{Henze} {et~al.}(2015{\natexlab{b}}){Henze}, {Ness}, {Darnley},
  {Bode}, {Williams}, {Shafter}, {Sala}, {Kato}, {Hachisu}, \&
  {Hernanz}}]{2015A&A...580A..46H}
{Henze}, M., {Ness}, J.-U., {Darnley}, M.~J., {et~al.} 2015{\natexlab{b}},
  \aap, 580, A46

\bibitem[{{Hillebrandt} \& {Niemeyer}(2000)}]{2000ARA&A..38..191H}
{Hillebrandt}, W., \& {Niemeyer}, J.~C. 2000, \araa, 38, 191

\bibitem[{{Hubble}(1929)}]{1929ApJ....69..103H}
{Hubble}, E.~P. 1929, \apj, 69, 103

\bibitem[{{Iijima}(2009)}]{2009A&A...505..287I}
{Iijima}, T. 2009, \aap, 505, 287

\bibitem[{{Jurdana-{\v S}epi{\'c}} {et~al.}(2012){Jurdana-{\v S}epi{\'c}},
  {Ribeiro}, {Darnley}, {Munari}, \& {Bode}}]{2012A&A...537A..34J}
{Jurdana-{\v S}epi{\'c}}, R., {Ribeiro}, V.~A.~R.~M., {Darnley}, M.~J.,
  {Munari}, U., \& {Bode}, M.~F. 2012, \aap, 537, A34

\bibitem[{{Kasliwal} {et~al.}(2008){Kasliwal}, {Quimby}, \&
  {Kulkarni}}]{2008ATel.1867....1K}
{Kasliwal}, M.~M., {Quimby}, R., \& {Kulkarni}, S.~R. 2008, The Astronomer's
  Telegram, 1867, 1

\bibitem[{{Kato} {et~al.}(2015){Kato}, {Saio}, \&
  {Hachisu}}]{2015ApJ...808...52K}
{Kato}, M., {Saio}, H., \& {Hachisu}, I. 2015, \apj, 808, 52

\bibitem[{{Kato} {et~al.}(2014){Kato}, {Saio}, {Hachisu}, \&
  {Nomoto}}]{2014ApJ...793..136K}
{Kato}, M., {Saio}, H., {Hachisu}, I., \& {Nomoto}, K. 2014, \apj, 793, 136

\bibitem[{{Kent}(1987)}]{1987AJ.....94..306K}
{Kent}, S.~M. 1987, \aj, 94, 306

\bibitem[{{Kolmogorov}(1933)}]{kolmogorov33}
{Kolmogorov}, A. 1933, Inst. Ital. Attuari., 4, 1

\bibitem[{{Law} {et~al.}(2009){Law}, {Kulkarni}, {Dekany}, {Ofek}, {Quimby},
  {Nugent}, {Surace}, {Grillmair}, {Bloom}, {Kasliwal}, {Bildsten}, {Brown},
  {Cenko}, {Ciardi}, {Croner}, {Djorgovski}, {van Eyken}, {Filippenko}, {Fox},
  {Gal-Yam}, {Hale}, {Hamam}, {Helou}, {Henning}, {Howell}, {Jacobsen},
  {Laher}, {Mattingly}, {McKenna}, {Pickles}, {Poznanski}, {Rahmer}, {Rau},
  {Rosing}, {Shara}, {Smith}, {Starr}, {Sullivan}, {Velur}, {Walters}, \&
  {Zolkower}}]{2009PASP..121.1395L}
{Law}, N.~M., {Kulkarni}, S.~R., {Dekany}, R.~G., {et~al.} 2009, \pasp, 121,
  1395

\bibitem[{{Maguire} {et~al.}(2013){Maguire}, {Sullivan}, {Patat}, {Gal-Yam},
  {Hook}, {Dhawan}, {Howell}, {Mazzali}, {Nugent}, {Pan}, {Podsiadlowski},
  {Simon}, {Sternberg}, {Valenti}, {Baltay}, {Bersier}, {Blagorodnova}, {Chen},
  {Ellman}, {Feindt}, {F{\"o}rster}, {Fraser}, {Gonz{\'a}lez-Gait{\'a}n},
  {Graham}, {Guti{\'e}rrez}, {Hachinger}, {Hadjiyska}, {Inserra}, {Knapic},
  {Laher}, {Leloudas}, {Margheim}, {McKinnon}, {Molinaro}, {Morrell}, {Ofek},
  {Rabinowitz}, {Rest}, {Sand}, {Smareglia}, {Smartt}, {Taddia}, {Walker},
  {Walton}, \& {Young}}]{2013MNRAS.436..222M}
{Maguire}, K., {Sullivan}, M., {Patat}, F., {et~al.} 2013, \mnras, 436, 222

\bibitem[{{Marquardt} {et~al.}(2015){Marquardt}, {Sim}, {Ruiter}, {Seitenzahl},
  {Ohlmann}, {Kromer}, {Pakmor}, \& {R{\"o}pke}}]{2015A&A...580A.118M}
{Marquardt}, K.~S., {Sim}, S.~A., {Ruiter}, A.~J., {et~al.} 2015, \aap, 580,
  A118

\bibitem[{{Montalto} {et~al.}(2009){Montalto}, {Seitz}, {Riffeser}, {Hopp},
  {Lee}, \& {Sch{\"o}nrich}}]{2009A&A...507..283M}
{Montalto}, M., {Seitz}, S., {Riffeser}, A., {et~al.} 2009, \aap, 507, 283

\bibitem[{{Munari} \& {Walter}(2016)}]{2016MNRAS.455L..57M}
{Munari}, U., \& {Walter}, F.~M. 2016, \mnras, 455, L57

\bibitem[{{Munari} {et~al.}(2011){Munari}, {Joshi}, {Ashok}, {Banerjee},
  {Valisa}, {Milani}, {Siviero}, {Dallaporta}, \&
  {Castellani}}]{2011MNRAS.410L..52M}
{Munari}, U., {Joshi}, V.~H., {Ashok}, N.~M., {et~al.} 2011, \mnras, 410, L52

\bibitem[{{Pagnotta} \& {Schaefer}(2014)}]{2014ApJ...788..164P}
{Pagnotta}, A., \& {Schaefer}, B.~E. 2014, \apj, 788, 164

\bibitem[{{Patat} {et~al.}(2007){Patat}, {Chandra}, {Chevalier}, {Justham},
  {Podsiadlowski}, {Wolf}, {Gal-Yam}, {Pasquini}, {Crawford}, {Mazzali},
  {Pauldrach}, {Nomoto}, {Benetti}, {Cappellaro}, {Elias-Rosa}, {Hillebrandt},
  {Leonard}, {Pastorello}, {Renzini}, {Sabbadin}, {Simon}, \&
  {Turatto}}]{2007Sci...317..924P}
{Patat}, F., {Chandra}, P., {Chevalier}, R., {et~al.} 2007, Science, 317, 924

\bibitem[{{Pietsch}(2010)}]{2010AN....331..187P}
{Pietsch}, W. 2010, Astronomische Nachrichten, 331, 187

\bibitem[{{Pietsch} {et~al.}(2007){Pietsch}, {Haberl}, {Sala}, {Stiele},
  {Hornoch}, {Riffeser}, {Fliri}, {Bender}, {B{\"u}hler}, {Burwitz}, {Greiner},
  \& {Seitz}}]{2007A&A...465..375P}
{Pietsch}, W., {Haberl}, F., {Sala}, G., {et~al.} 2007, \aap, 465, 375

\bibitem[{{Pottasch}(1967)}]{1967BAN....19..227P}
{Pottasch}, S.~R. 1967, \bain, 19, 227

\bibitem[{{Schaefer}(2010)}]{2010ApJS..187..275S}
{Schaefer}, B.~E. 2010, \apjs, 187, 275

\bibitem[{{Shafter} {et~al.}(2011{\natexlab{a}}){Shafter}, {Bode}, {Darnley},
  {Misselt}, {Rubin}, \& {Hornoch}}]{2011ApJ...727...50S}
{Shafter}, A.~W., {Bode}, M.~F., {Darnley}, M.~J., {et~al.} 2011{\natexlab{a}},
  \apj, 727, 50

\bibitem[{{Shafter} {et~al.}(2012){Shafter}, {Darnley}, {Bode}, \&
  {Ciardullo}}]{2012ApJ...752..156S}
{Shafter}, A.~W., {Darnley}, M.~J., {Bode}, M.~F., \& {Ciardullo}, R. 2012,
  \apj, 752, 156

\bibitem[{{Shafter} \& {Irby}(2001)}]{2001ApJ...563..749S}
{Shafter}, A.~W., \& {Irby}, B.~K. 2001, \apj, 563, 749

\bibitem[{{Shafter} {et~al.}(2009){Shafter}, {Rau}, {Quimby}, {Kasliwal},
  {Bode}, {Darnley}, \& {Misselt}}]{2009ApJ...690.1148S}
{Shafter}, A.~W., {Rau}, A., {Quimby}, R.~M., {et~al.} 2009, \apj, 690, 1148

\bibitem[{{Shafter} {et~al.}(2011{\natexlab{b}}){Shafter}, {Darnley},
  {Hornoch}, {Filippenko}, {Bode}, {Ciardullo}, {Misselt}, {Hounsell},
  {Chornock}, \& {Matheson}}]{2011ApJ...734...12S}
{Shafter}, A.~W., {Darnley}, M.~J., {Hornoch}, K., {et~al.} 2011{\natexlab{b}},
  \apj, 734, 12

\bibitem[{{Shafter} {et~al.}(2015){Shafter}, {Henze}, {Rector}, {Schweizer},
  {Hornoch}, {Orio}, {Pietsch}, {Darnley}, {Williams}, {Bode}, \&
  {Bryan}}]{2015ApJS..216...34S}
{Shafter}, A.~W., {Henze}, M., {Rector}, T.~A., {et~al.} 2015, \apjs, 216, 34

\bibitem[{{Shore} {et~al.}(2011){Shore}, {Wahlgren}, {Augusteijn}, {Liimets},
  {Page}, {Osborne}, {Beardmore}, {Koubsky}, {{\v S}lechta}, \&
  {Votruba}}]{2011A&A...527A..98S}
{Shore}, S.~N., {Wahlgren}, G.~M., {Augusteijn}, T., {et~al.} 2011, \aap, 527,
  A98

\bibitem[{{Silverman} {et~al.}(2013){Silverman}, {Nugent}, {Gal-Yam},
  {Sullivan}, {Howell}, {Filippenko}, {Arcavi}, {Ben-Ami}, {Bloom}, {Cenko},
  {Cao}, {Chornock}, {Clubb}, {Coil}, {Foley}, {Graham}, {Griffith}, {Horesh},
  {Kasliwal}, {Kulkarni}, {Leonard}, {Li}, {Matheson}, {Miller}, {Modjaz},
  {Ofek}, {Pan}, {Perley}, {Poznanski}, {Quimby}, {Steele}, {Sternberg}, {Xu},
  \& {Yaron}}]{2013ApJS..207....3S}
{Silverman}, J.~M., {Nugent}, P.~E., {Gal-Yam}, A., {et~al.} 2013, \apjs, 207,
  3

\bibitem[{{Simon} {et~al.}(2009){Simon}, {Gal-Yam}, {Gnat}, {Quimby},
  {Ganeshalingam}, {Silverman}, {Blondin}, {Li}, {Filippenko}, {Wheeler},
  {Kirshner}, {Patat}, {Nugent}, {Foley}, {Vogt}, {Butler}, {Peek},
  {Rosolowsky}, {Herczeg}, {Sauer}, \& {Mazzali}}]{2009ApJ...702.1157S}
{Simon}, J.~D., {Gal-Yam}, A., {Gnat}, O., {et~al.} 2009, \apj, 702, 1157

\bibitem[{{Smirnov}(1948)}]{smirnov1948}
{Smirnov}, N. 1948, Ann. Math. Statist., 19, 279

\bibitem[{{Soker} {et~al.}(2013){Soker}, {Kashi}, {Garc{\'{\i}}a-Berro},
  {Torres}, \& {Camacho}}]{2013MNRAS.431.1541S}
{Soker}, N., {Kashi}, A., {Garc{\'{\i}}a-Berro}, E., {Torres}, S., \&
  {Camacho}, J. 2013, \mnras, 431, 1541

\bibitem[{{Stark} {et~al.}(1992){Stark}, {Gammie}, {Wilson}, {Bally}, {Linke},
  {Heiles}, \& {Hurwitz}}]{1992ApJS...79...77S}
{Stark}, A.~A., {Gammie}, C.~F., {Wilson}, R.~W., {et~al.} 1992, \apjs, 79, 77

\bibitem[{{Starrfield} {et~al.}(2012){Starrfield}, {Iliadis}, {Timmes}, {Hix},
  {Arnett}, {Meakin}, \& {Sparks}}]{2012BASI...40..419S}
{Starrfield}, S., {Iliadis}, C., {Timmes}, F.~X., {et~al.} 2012, Bulletin of
  the Astronomical Society of India, 40, 419

\bibitem[{{Starrfield} {et~al.}(1972){Starrfield}, {Truran}, {Sparks}, \&
  {Kutter}}]{1972ApJ...176..169S}
{Starrfield}, S., {Truran}, J.~W., {Sparks}, W.~M., \& {Kutter}, G.~S. 1972,
  \apj, 176, 169

\bibitem[{{Steele} {et~al.}(2004){Steele}, {Smith}, {Rees}, {Baker}, {Bates},
  {Bode}, {Bowman}, {Carter}, {Etherton}, {Ford}, {Fraser}, {Gomboc}, {Lett},
  {Mansfield}, {Marchant}, {Medrano-Cerda}, {Mottram}, {Raback}, {Scott},
  {Tomlinson}, \& {Zamanov}}]{2004SPIE.5489..679S}
{Steele}, I.~A., {Smith}, R.~J., {Rees}, P.~C., {et~al.} 2004, in Society of
  Photo-Optical Instrumentation Engineers (SPIE) Conference Series, Vol. 5489,
  Ground-based Telescopes, ed. J.~M. {Oschmann}, Jr., 679--692

\bibitem[{{Stephens}(1974)}]{doi:10.1080/01621459.1974.10480196}
{Stephens}, M.~A. 1974, Journal of the American Statistical Association, 69,
  730

\bibitem[{{Sternberg} {et~al.}(2011){Sternberg}, {Gal-Yam}, {Simon}, {Leonard},
  {Quimby}, {Phillips}, {Morrell}, {Thompson}, {Ivans}, {Marshall},
  {Filippenko}, {Marcy}, {Bloom}, {Patat}, {Foley}, {Yong}, {Penprase},
  {Beeler}, {Allende Prieto}, \& {Stringfellow}}]{2011Sci...333..856S}
{Sternberg}, A., {Gal-Yam}, A., {Simon}, J.~D., {et~al.} 2011, Science, 333,
  856

\bibitem[{{Strope} {et~al.}(2010){Strope}, {Schaefer}, \&
  {Henden}}]{2010AJ....140...34S}
{Strope}, R.~J., {Schaefer}, B.~E., \& {Henden}, A.~A. 2010, \aj, 140, 34

\bibitem[{{Tang} {et~al.}(2014){Tang}, {Bildsten}, {Wolf}, {Li}, {Kong}, {Cao},
  {Cenko}, {De Cia}, {Kasliwal}, {Kulkarni}, {Laher}, {Masci}, {Nugent},
  {Perley}, {Prince}, \& {Surace}}]{2014ApJ...786...61T}
{Tang}, S., {Bildsten}, L., {Wolf}, W.~M., {et~al.} 2014, \apj, 786, 61

\bibitem[{{Williams} \& {Mason}(2010)}]{2010Ap&SS.327..207W}
{Williams}, R., \& {Mason}, E. 2010, \apss, 327, 207

\bibitem[{{Williams}(2014)}]{williamsPhD}
{Williams}, S.~C. 2014, PhD thesis, Astrophysics Research Institute, Liverpool
  John Moores, University, IC2 Liverpool Science Park, 146 Brownlow Hill,
  Liverpool, L3 5RF, UK

\bibitem[{{Williams} {et~al.}(2014){Williams}, {Darnley}, {Bode}, {Keen}, \&
  {Shafter}}]{2014ApJS..213...10W}
{Williams}, S.~C., {Darnley}, M.~J., {Bode}, M.~F., {Keen}, A., \& {Shafter},
  A.~W. 2014, \apjs, 213, 10 (Paper~I)

\bibitem[{{Williams} {et~al.}(2015){Williams}, {Darnley}, {Bode}, \&
  {Steele}}]{2015ApJ...805L..18W}
{Williams}, S.~C., {Darnley}, M.~J., {Bode}, M.~F., \& {Steele}, I.~A. 2015,
  \apjl, 805, L18

\bibitem[{{Woudt} \& {Ribeiro}(2014)}]{2014ASPC..490.....W}
{Woudt}, P.~A., \& {Ribeiro}, V.~A.~R.~M., eds. 2014, Astronomical Society of
  the Pacific Conference Series, Vol. 490, {Stella Novae: Past and Future
  Decades} (San Francisco: Astronomical Society of the Pacific)

\end{thebibliography}

\end{document}